\documentclass[aps,prb,reprint,groupedaddress]{revtex4-1}

\usepackage{float}
\usepackage{amsmath}
\usepackage{amssymb}
\usepackage{bm}
\usepackage{graphicx}
\usepackage{color}
\usepackage{hyperref}
\hypersetup{colorlinks,citecolor=blue,linkcolor=blue,urlcolor=blue}

\definecolor{myred}{RGB}{170,0,0}

\begin{document}

\title{Inverse design of disordered stealthy hyperuniform spin chains}

\author{Eli Chertkov}
\affiliation{Department of Physics, Princeton University, Princeton, New Jersey 08544, USA}
\author{Robert A. DiStasio Jr.}
\affiliation{Department of Chemistry, Princeton University, Princeton, New Jersey 08544, USA}
\affiliation{Department of Chemistry and Chemical Biology, Cornell University, Ithaca, New York 14853, USA}
\author{Ge Zhang}
\affiliation{Department of Chemistry, Princeton University, Princeton, New Jersey 08544, USA}
\author{Roberto Car}
\author{Salvatore Torquato}
\affiliation{Department of Chemistry, Princeton University, Princeton, New Jersey 08544, USA}
\affiliation{Department of Physics, Princeton University, Princeton, New Jersey 08544, USA}
\affiliation{Princeton Institute for the Science and Technology of Materials, Princeton University, Princeton, New Jersey 08544, USA}
\affiliation{Program in Applied and Computational Mathematics, Princeton University, Princeton, New Jersey 08544, USA}

\date{\today}

\begin{abstract}
Positioned between crystalline solids and liquids, disordered many-particle systems which are stealthy and hyperuniform represent new states of matter that are endowed with novel physical and thermodynamic properties. Such stealthy and hyperuniform states are unique in that they are transparent to radiation for a range of wavenumbers around the origin.
In this work, we employ recently developed inverse statistical-mechanical methods, which seek to obtain the optimal set of interactions that will spontaneously produce a targeted structure or configuration as a unique ground state, to investigate the spin-spin interaction potentials required to stabilize disordered stealthy hyperuniform one-dimensional (1D) Ising-like spin chains.
By performing an exhaustive search over the spin configurations that can be enumerated on periodic 1D integer lattices containing $N=2,3,\ldots,36$ sites, we were able to identify and structurally characterize \textit{all} stealthy hyperuniform spin chains in this range of system sizes.
Within this pool of stealthy hyperuniform spin configurations, we then utilized such inverse optimization techniques to demonstrate that stealthy hyperuniform spin chains can be realized as either unique or degenerate disordered ground states of radial long-ranged (relative to the spin chain length) spin-spin interactions.
Such exotic ground states are distinctly different from spin glasses in both their inherent structural properties and the nature of the spin-spin interactions required to stabilize them.
As such, the implications and significance of the existence of such disordered stealthy hyperuniform ground state spin systems warrants further study, including whether their bulk physical properties and excited states, like their many-particle system counterparts, are singularly remarkable, and can be experimentally realized.
\end{abstract}

\maketitle

\section{Introduction}

There has been a tremendous amount of interest recently in the creation of ``materials by design,'' that is, the directed and systematic search for new materials that possess prescribed desirable properties. Computational methods developed for this purpose will play a central role toward this goal. Inverse statistical-mechanical methods allow for a new mode of thinking about the structure and physical properties of condensed phases of matter,~\cite{inverseopt} and are ideally suited for materials discovery by design. 

Much of statistical mechanics centers around finding the structural and bulk physical properties for a given many-particle (or many-spin) system Hamiltonian, what we refer to as the ``forward'' problem of statistical mechanics.~\cite{ising,potts,onsager,yang} In this regard, so-called inverse statistical-mechanical methods have been devised that yield optimized interactions that robustly and spontaneously lead to a targeted many-particle configuration of the system or targeted set of physical properties for a wide range of conditions.~\cite{inverseopt,torquato_honeycomb,torquato_triangularsquarehoneycomb,cohnkumar,torquato_monotonicsquarehoneycomb,torquato_diamond,muller,hannon,zhang,jain1,jain2,negativethermalexpansion,negativepoissonratio,zunger1,zunger2,zunger3} An interesting class of target configurations that have been studied are classical many-particle ground states of varying complexity and novelty.~\cite{inverseopt,torquato_honeycomb,torquato_triangularsquarehoneycomb,cohnkumar,torquato_monotonicsquarehoneycomb,torquato_diamond,muller,hannon,zhang,jain1,jain2} One can also design the system to have exotic physical properties, such a negative Poisson ratio and negative thermal expansion coefficients over a range of temperatures.~\cite{negativethermalexpansion,negativepoissonratio}

Recently, we have generalized inverse statistical-mechanical methods to the case of two-state Ising spin systems with radial spin-spin interactions of finite range (i.e., extending beyond nearest-neighbor sites).~\cite{designer1,designer2} Our interest in these initial studies was to find the optimal set of shortest-range pair interactions whose corresponding ground state was a targeted spin configuration. The possible outcomes for a given target configuration were classified into whether or not the targeted ground-state spin configurations were unique, degenerate, or neither. In general, inverse techniques applied to targeted spin configurations could have implications for the design of solids with novel magnetic and electronic properties, \textit{e.g.} see Refs.~\onlinecite{designer1,stoner}.

Our objective in this work is to determine whether one-dimensional (1D) spin chains having target configurations that are disordered, stealthy, and hyperuniform can be made to be classical ground states. This choice is motivated by the fact that such exotic amorphous states of matter in the context of particle systems offer fascinating open theoretical questions and possess novel physical properties. Hyperuniform many-particle systems possess anomalously suppressed infinite-wavelength density fluctuations, as quantified by the number variance.~\cite{2003_torquato,hyperuniformpointpatterns} Disordered hyperuniformity  occurs in a variety of physical systems, including large-scale structure in the Universe,~\cite{statisticalphysicscosmic,peebles} the arrangement of avian photoreceptors,~\cite{avianphotoreceptors} driven non-equilibrium systems,~\cite{hexner,jack} dynamics in cold atoms,~\cite{lesanovsky} surface-enhanced Raman spectroscopy,~\cite{ramanspectroscopy} terahertz quantum cascade lasers,~\cite{quantumcascadelaser} wave dynamics in disordered potentials based on super-symmetry,~\cite{blochwavedynamics} and certain Coulombic systems.~\cite{pointprocesses}

Stealthy hyperuniform systems are those that completely suppresses single scattering for a range of long wavelengths.~\cite{stealthydisorder,torquato_arxiv} It has been shown that systems of particles interacting with certain long-ranged (\textit{i.e.}, on the order of the chain length) pair potentials can counterintuitively freeze into classical ground states that are disordered stealthy hyperuniform. By mapping such stealthy configurations of particles into network solids, the first disordered cellular solids with complete isotropic photonic band gaps comparable in size to photonic crystals were discovered.~\cite{2009_torquato,isotropicbandgaps}

Here we begin a program to apply inverse techniques to explore whether 1D spin chains that are disordered, stealthy and hyperuniform can be made to be classical ground states of radial spin-spin interactions. Such spin chains would suppress single scattering for a range of wavenumbers around the origin, implying anomalously suppressed magnetization fluctations at long wavelengths. By focusing only on configurations that are stealthy at the smallest positive wavenumber,  we are able to enumerate all periodic stealthy spin configurations containing $N=2,3,\ldots,36$ spins on the 1D integer lattice under periodic boundary conditions and then structurally characterize these configurations by computing pair correlation functions, structure factors, and degree of disorder. We select from them the disordered configurations and determine via inverse optimization techniques whether they can be made to be ground states. We discover that stealthy hyperuniform spin configurations can be realized as either unique or degenerate disordered ground states of radial long-ranged (\textit{i.e.}, relative to the chain length) spin-spin interactions. Such exotic ground states are distinctly different from spin glasses~\cite{spinglasses,parisi,nishimori} in both their inherent structural properties and the nature of the spin-spin interactions required to stabilize them (\textit{e.g.}, these disordered stealthy hyperuniform spin systems exhibit no single-scattering for large wavelengths and are stabilized by deterministic rather than stochastic spin-spin interactions).

Section~\ref{sec:theory} briefly reviews basic concepts, including definitions of stealthiness, hyperuniformity, and a disorder metric for 1D spin chains. Section~\ref{sec:methods} describes the inverse statistical mechanics procedure used to generate stabilizing radial spin-spin interactions. Section~\ref{sec:results} presents our major results, including an enumeration of stealthy hyperuniform spin configurations containing $N=2,3,\ldots,36$ spins on the 1D integer lattice under periodic boundary conditions, their characteristics, and their ability to be spontaneously generated as ground states by the inverse methodology. Finally, in Sec.~\ref{sec:conclusions}, we close with concluding remarks.

\section{Theoretical background} \label{sec:theory}

We are interested in a generalized class of spin Hamiltonians in arbitrary Euclidean space dimensions. In this initial study, we focus on the spin-spin interaction Hamiltonian of spin chains on the 1D integer lattice given by the distance-dependent version of the Ising model
\begin{align} 
H = -\sum_{R}J(R)\sigma_i\sigma_{i+R}
\end{align}
where $-1 \leq J(R) \leq 1$ can in general be a long-ranged spin-spin interaction potential with a radial extent on the order of the chain length ($R \approx N$). Such long-ranged spin-spin interactions have been studied by various investigators using standard \textit{forward} statistical-mechanical techniques.~\cite{ising,introANNNImodel,ANNNImodel,reviewANNNImodel,dysonSpinChain,1Dspinglass,lujiten,parisiLRspinchain} The inverse statistical mechanics methodology involves optimally tuning the $J(R)$ parameters such that the targeted spin configuration is spontaneously produced as a unique ground state according to the criteria and procedures described in Sec.~\ref{sec:methods}. In what follows, we briefly review the basic concepts and definitions that will be used throughout the paper, including collective coordinates, hyperuniformity and stealthiness, and order metrics.

\subsection{Collective coordinates}

The configurations we designate as targets in the inverse methodology are defined through collective density coordinates. Specifically, we focus on 1D spin chains composed of $N$ spins $\sigma_j=\pm 1$ positioned on the sites of the integer lattice $R_j = j$ in a fundamental unit cell under periodic boundary conditions. The spin collective density variable of the chains is the Fourier transform of the spin density
\begin{align}
\rho_\sigma(k)=\sum_{j=1}^N \sigma_j e^{ikj}, \label{eq:rho}
\end{align}
which can be viewed as a transformation from the finite set of spin configurations to the complex functions $\rho_\sigma(k)$ that depends on the infinite set of wavenumbers $k$ on the integer lattice in reciprocal space.~\cite{torquato_arxiv} The possible wavevectors for a chain on the integer lattice are $k=2\pi n/N$ for integer $n$. The spin structure factor is then defined as
\begin{align} 
S(k) = \frac{1}{N}\rho_\sigma(k)\rho_\sigma(-k), \label{eq:Sk}
\end{align}
which is a real function with inversion symmetry. For spins on the integer lattice, the structure factor is periodic with $S(k)=S(k+2\pi)$. This symmetry property combined with Eq. (\ref{eq:Sk}) imply that $S(k)$ is symmetric about $k=\pi$ for integer spin chains. In addition, through Fourier inversion of the structure factor we can obtain the radial spin-spin correlation function
\begin{align}
S_2(r) = \frac{1}{N}\sum_{j=1}^N \sigma_j \sigma_{j+r}. \label{eq:S2}
\end{align}

We note here that the structure factor is expressible in terms of the spin-spin correlation function:
\[ S(k) = \sum_{r=1}^NS_2(r)e^{ikr}. \]

\subsection{Hyperuniformity and stealthiness}

The spin configurations of interest in this study possess both hyperuniformity and stealthiness. Hyperuniform spin systems possess anomalously suppressed infinite-wavelength magnetization fluctuations, that is~\cite{2003_torquato}
\[ \lim_{k\rightarrow 0}S(k) = 0, \]
a definition that explicitly omits forward scattering.

All periodic patterns are hyperuniform, but disordered configurations, such as liquids and structural glasses, generally are not. It is well known that the $k \rightarrow 0$ limit of the structure factor for particle systems in thermal equilibrium is proportional to the isothermal compressibility.~\cite{simpleliquids} 

Stealthy configurations are those in which the structure factor is zero for a range of wavevectors. In the special case in which $S(k)=0$ for $0< k \le K$ (where $K>0$ is called the exclusion zone radius), the stealthy configurations are also hyperuniform, which is the case of interest in the present paper. Stealthy patterns are characterized by long-ranged correlations due to the suppression of large-scale density fluctuations. In this sense, disordered stealthy hyperuniform patterns have a ``hidden order'' on large length scales.~\cite{torquato_arxiv} There has been a recent effort to generate stealthy particle configurations through ``stealthy'' potentials; see Ref.~\onlinecite{torquato_arxiv} and references therein. These stealthy potentials are radial pairwise interaction potentials restricted to a range of wavevectors. These potentials have been shown to produce highly degenerate disordered, hyperuniform, stealthy configurations.~\cite{torquato_arxiv} 

\begin{figure}[t]
\begin{center}
\begin{tabular}{c}
\includegraphics[width=0.5\textwidth]{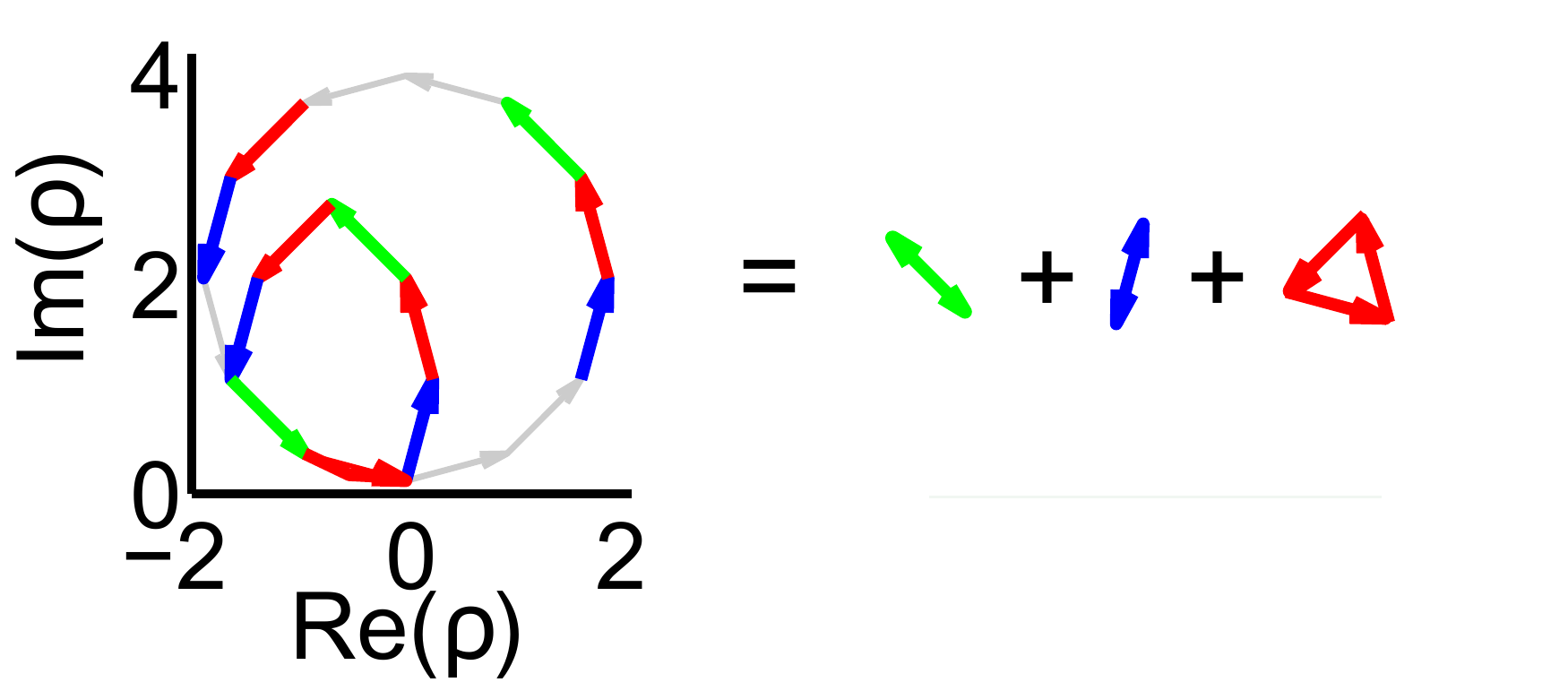} \\
(a) \vspace{0.2cm} \\
\vspace{0.2cm}
\includegraphics[width=0.25\textwidth]{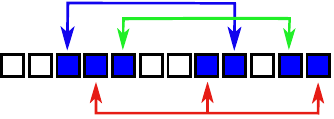} \\
(b)
\end{tabular}
\end{center}
\caption{(a) A visualization of how $\rho(k)$ cancels at the smallest positive wavenumber $k=2\pi/N$ for (b) the depicted stealthy spin configuration of size $N=12$ on the 1D integer lattice. The unit vectors represent the positive spin ($\sigma_j=+1$) exponential $e^{i2\pi j/N}$ terms from the collective density variables $\rho(k=2\pi/N)=\sum_{j=1}^N\sigma_je^{i2\pi j/N}$. The negative spin terms also cancel in a similar fashion. The plot contains a gray regular polygon including all possible unit vectors. The unit vectors that are summed for the positive spins in this particular configuration are colored according to how they cancel with other vectors. In this case, the scattering cancellation can be decomposed into two doublets (green and blue) and one triplet (red), which correspond to the indicated spins in (b).} \label{fig:M12_polygon}
\end{figure}

In this study, we explore the discrete spin chain analogs of these continuous stealthy hyperuniform particle systems via the inverse statistical mechanics methodology. Specifically, we investigate periodic 1D spin chains on the integer lattice that are both stealthy and hyperuniform, so that $S(k)=0$ for at least the range of wavenumbers $0 < k \leq K= 2\pi/N$. Obtaining stealthy spin chains for $K>0$ can be non-trivial, as demonstrated in Fig.~\ref{fig:M12_polygon}. The complex exponentials in the collective density variables [cf. Eq. (\ref{eq:rho})] must cancel at least at the smallest positive wavenumber $k=2\pi/N$ for the spin configuration in question to be considered stealthy hyperuniform.

\subsection{Order metric}

We are particularly interested in stealthy hyperuniform configurations that possess a high degree of disorder. As discussed in the Appendix, periodic spin configurations can appear ``reducible.'' For reducible stealthy spin patterns, the unit cell of the pattern can be expressed as a repetition of a smaller unit cell. These patterns will of course appear to be \emph{more ordered}. Of greater interest to us are those spin patterns that cannot be decomposed into smaller unit cells: the irreducible patterns. These patterns are more disordered.

We quantify the degree of order between configurations of the same size $N$ and magnetization $\langle\sigma\rangle = \sum_{j=1}^N\sigma_j/N$ by using the order metric defined by Ref.~\onlinecite{torquato_arxiv} and used for two-dimensional stealthy spin systems in Ref.~\onlinecite{robpaper}:
\begin{align}
\tau = \sum_{k} \left[ S(k) - S_0(k) \right]^2 \label{eq:tau}
\end{align}
where the summation is over $k=2\pi/N,4\pi/N,\ldots,2\pi(N-1)/N$ and $S_0(k)$ is a reference structure factor. For our purposes, we set this to a constant value
\begin{align} 
S_0(k) = 1 - \langle\sigma\rangle^2,
\end{align}
which is the expected structure factor of an ensemble of uncorrelated Poisson spin patterns of period $N$ and average magnetization $\langle \sigma \rangle$. Each configuration in the Poisson ensemble is generated by setting each spin to $\sigma_i=+1$ with probability $\langle\sigma\rangle$ and to $\sigma_i=-1$ otherwise.

\begin{figure}[t]
\begin{center}
\begin{tabular}{|c|}
\hline
{\large (a) {\color{myred}$\tau=869$}} \\
\includegraphics[width=0.43\textwidth]{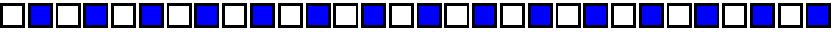} \\
\includegraphics[width=0.43\textwidth]{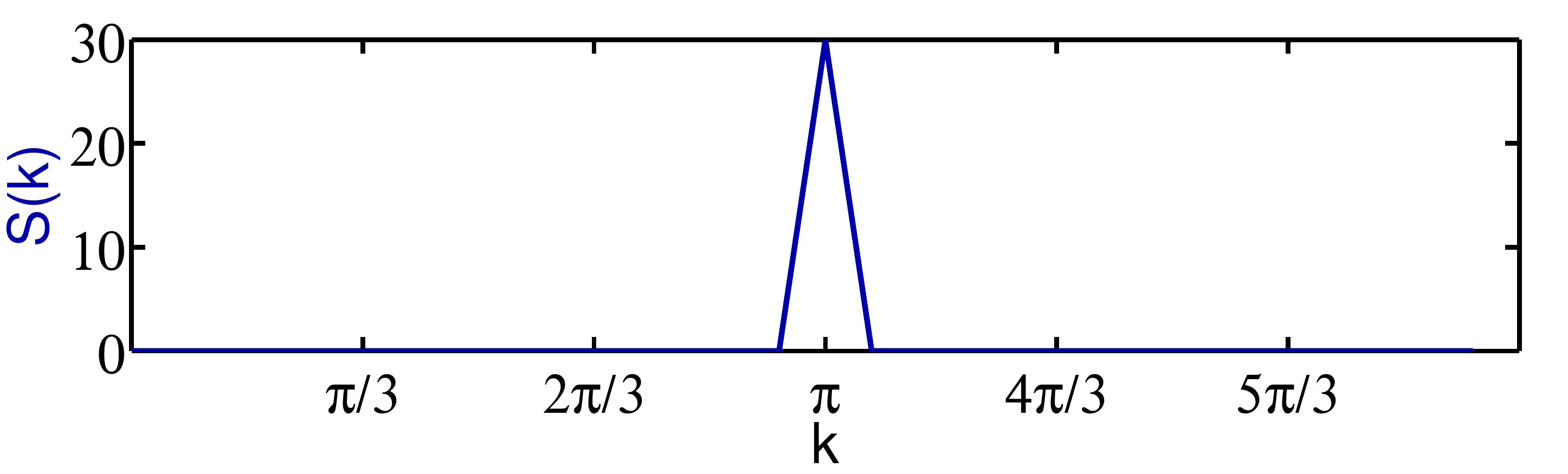} \\
\hline
{\large (b) {\color{myred} $\tau=103.1$}} \\
\includegraphics[width=0.43\textwidth]{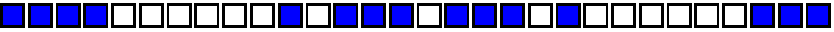} \\
\includegraphics[width=0.43\textwidth]{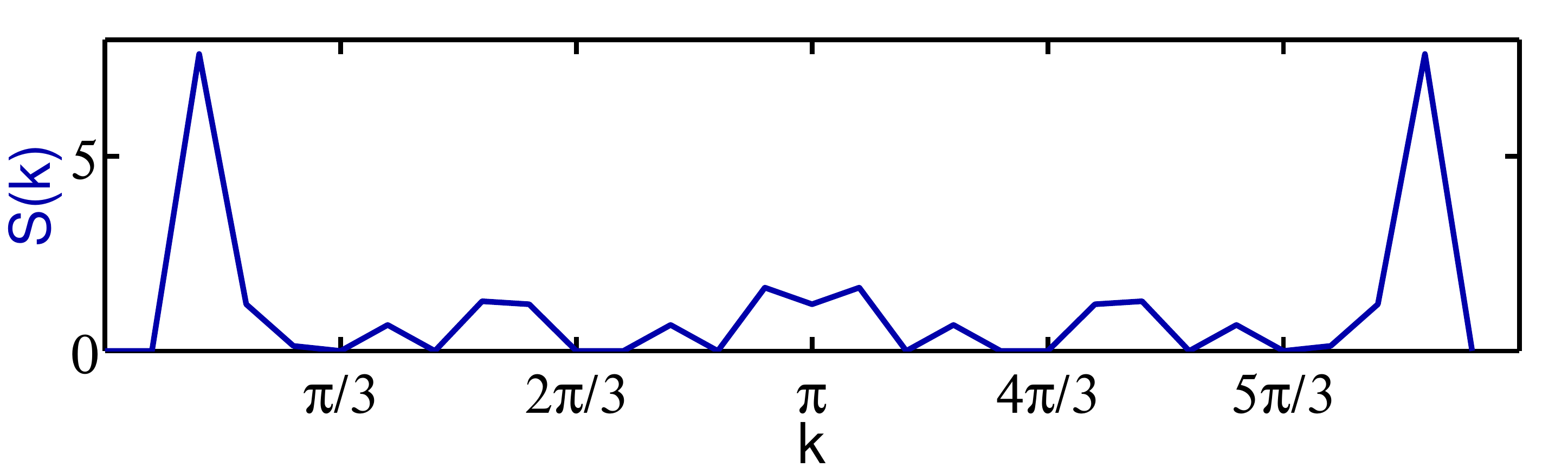} \\
\hline
{\large (c) {\color{myred} $\tau=30.6$}} \\
\includegraphics[width=0.43\textwidth]{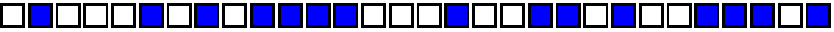} \\
\includegraphics[width=0.43\textwidth]{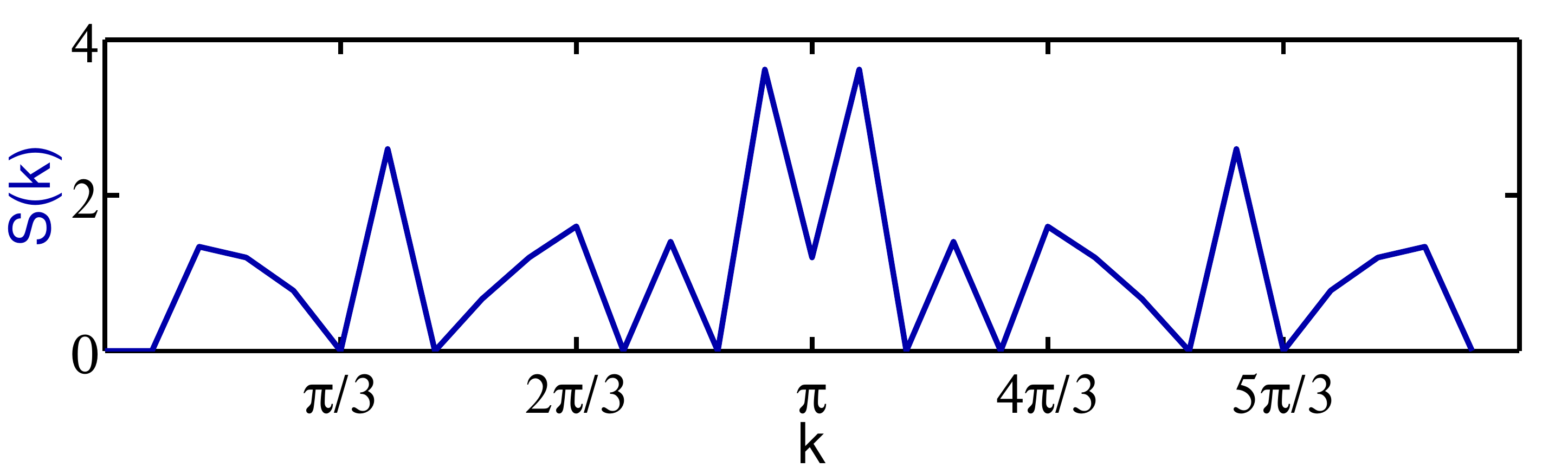} \\
\hline \hline
{ \large (d) {\color{myred} $\tau=13.5$}} \\
\includegraphics[width=0.43\textwidth]{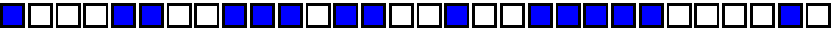} \\
\includegraphics[width=0.43\textwidth]{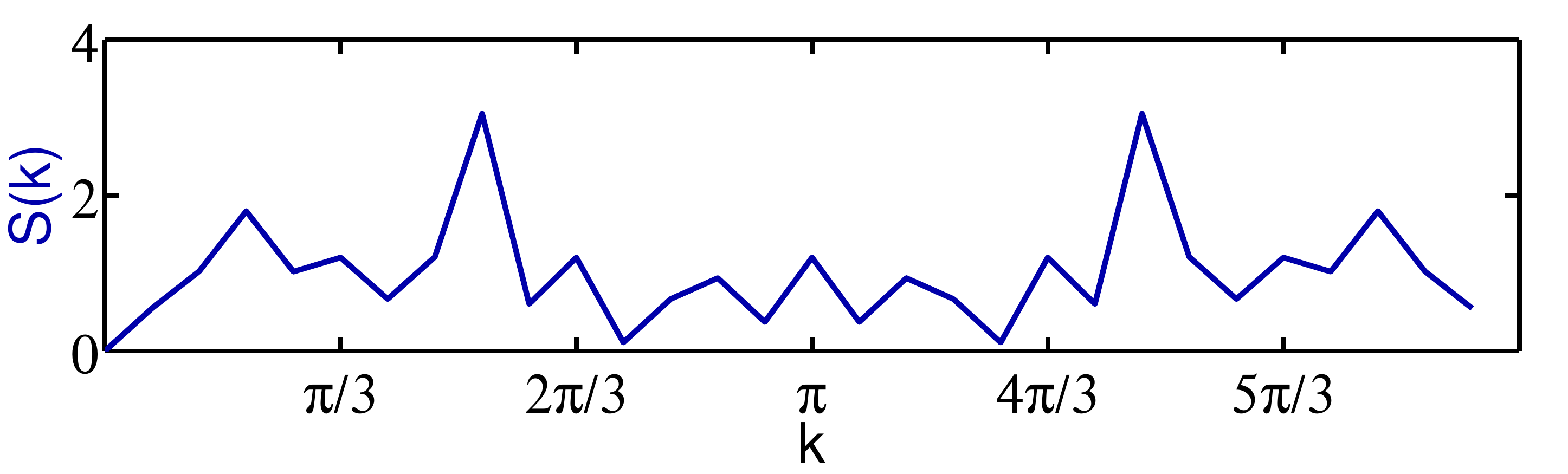} \\
\hline
\end{tabular}
\end{center}
\caption{Four periodic spin configurations with $N=30$ sites and $\langle\sigma
\rangle = 0$. The order metric $\tau$ of each configuration is displayed, followed by the spin configuration, and the structure factor. (a)--(c) The top three are stealthy hyperuniform. (d) The bottom configuration is a randomly generated Poisson spin pattern, which is hyperuniform but not stealthy.} \label{fig:confs_order}
\end{figure}

The order metric (\ref{eq:tau}) is meaningful for relative comparisons between spin configurations of fixed size $N$ and fixed magnetization $\langle\sigma\rangle$. Four examples of spin configurations with $\langle\sigma\rangle=0$ and different $\tau$ are displayed in Fig.~\ref{fig:confs_order}(a)--(d) along with their corresponding structure factors. Configurations (a)--(c) are stealthy hyperuniform, while (d) is a random Poisson pattern. As one can see, configuration (a) is periodic on a unit cell of size 2 and has a single Bragg peak in $0<k<2\pi$. This large deviation of $S(k)$ from $S_0=1$ at $k=\pi$ leads to a large $\tau$ value. Configuration (b), the second most ordered configuration, is almost reflectionally symmetric about the center of its unit cell, which leads to two peaks smaller in size than in (a). This explains the elevated $\tau$ value. The remaining two configurations (c) and (d) do not have any easily discernible order and both have $S(k)$ vary about the target structure factor $S_0=1$. Yet, we see that the order metric can pick up the ``hidden order'' of (c) the stealthy hyperuniform configuration compared to (d) the Poisson pattern.

We note in passing that a direct-space representation of the order metric defined in (\ref{eq:tau}) is trivially obtained by Parseval's theorem,~\cite{torquato_arxiv} which is effectively tantamount to replacing $S(k)$ with the direct-space two-point function $S_2(r)$ [defined by (\ref{eq:S2})] and summing over $r$. Such direct-space sums have been studied in a variety of other contexts (not as order metrics), including spin Hamiltonians~\cite{parisibinary} and as optimization objective functionals to reconstruct digitized binary heterogeneous media.~\cite{reconstruction}

\section{Inverse optimization techniques} \label{sec:methods}

We systematically study all stealthy hyperuniform spin configurations under periodic boundary conditions with $N=2,\ldots,36$. Specifically, we enumerate the $2^N$ possible configurations and keep only those with a vanishing structure factor for at least the smallest positive wavenumber $k=2\pi/N$. In our enumeration, two configurations that can be produced by translations or reflections of each other are considered equivalent and are not counted as distinct configurations.~\cite{designer1,designer2}

We characterize the structural properties of the discovered stealthy spin systems, including their degree of disorder, and attempt to generate spin-spin interaction potentials, which produce the stealthy spin configurations as ground states if such potentials exist. We accomplish the latter via the inverse statistical mechanics procedure for spin systems developed in Ref.~\onlinecite{designer1}. The procedure searches for interactions $J(R)$ in two-state spin Hamiltonians of the form
\begin{align}
H &= -\sum_{i=1}^N\sum_{R=1}^{R_C}J(R)\sigma_i\sigma_{i+R} \notag \\
 &= -N\sum_{R=1}^{R_C}J(R)S_2(R) \label{eq:H}
\end{align}
with interactions up to a cutoff distance $R_C$ between spins. The procedure attempts to produce a target configuration $\mathcal{T}$ as a ground state by maximizing the difference in energy density ($\epsilon = H/N$) between $\mathcal{T}$ and the closest energetic competitor $C$:
\[ \Delta\epsilon^C = \epsilon^C-\epsilon^\mathcal{T} = -\sum_{R=1}^{R_C}J(R)\left[S_2^C(R) - S_2^\mathcal{T}(R) \right] \]
subject to the constraint that $\Delta\epsilon^C \geq 0$ and $-1 \leq J(R) \leq 1$. Note that $R_C \leq R_{max}$ must hold, for a certain maximum distance $R_{max}$, so that the inverse problem is feasible.

There are three possible outcomes for the inverse method, which can be organized into the following solution classes:
\begin{itemize}
\item Class I: Solutions in which a spin-spin potential $J(R)$ is found that generates the target configuration as the unique ground state up to translations, reflections, and spin inversion operations.
\item Class II: Solutions in which a spin-spin potential is found that generates the target configuration as a non-unique ground state, degenerate in spin-spin correlation $S_2(R)$ [cf. Eq.~(\ref{eq:S2})] with other spin configurations.
\item Class III: Solutions that are neither class I nor II.
\end{itemize}
For a given competitor list, this optimization problem is solved using linear programming (to within numerical precision) and then simulated annealing with Metropolis Markov Chain Monte Carlo sampling is used to generate additional competitor spin configurations. Note that Class III solutions only arise when the optimization is deemed unfeasible by the linear programming methods. As such, there are no false-positive Class III solutions. For more details on the inverse statistical mechanics method applied to spin systems, see Refs.~\onlinecite{designer1,designer2}.

\section{Results and Discussion} \label{sec:results}

With a complete list of stealthy spin configurations of size $N=2,\ldots,36$ at our disposal, we now examine their statistical properties, such as magnetization and degree of disorder, and classify their ``groundstateability,'' i.e., whether such target spin configurations can be stabilized as unique or degenerate ground states.

\vspace{4mm}
\subsection{Characterization of enumerated configurations}

\subsubsection{Spin configurations of fixed size and magnetization}

\begin{figure*}[t]
\begin{center}
\begin{tabular}{cc}
\includegraphics[width=0.4\textwidth]{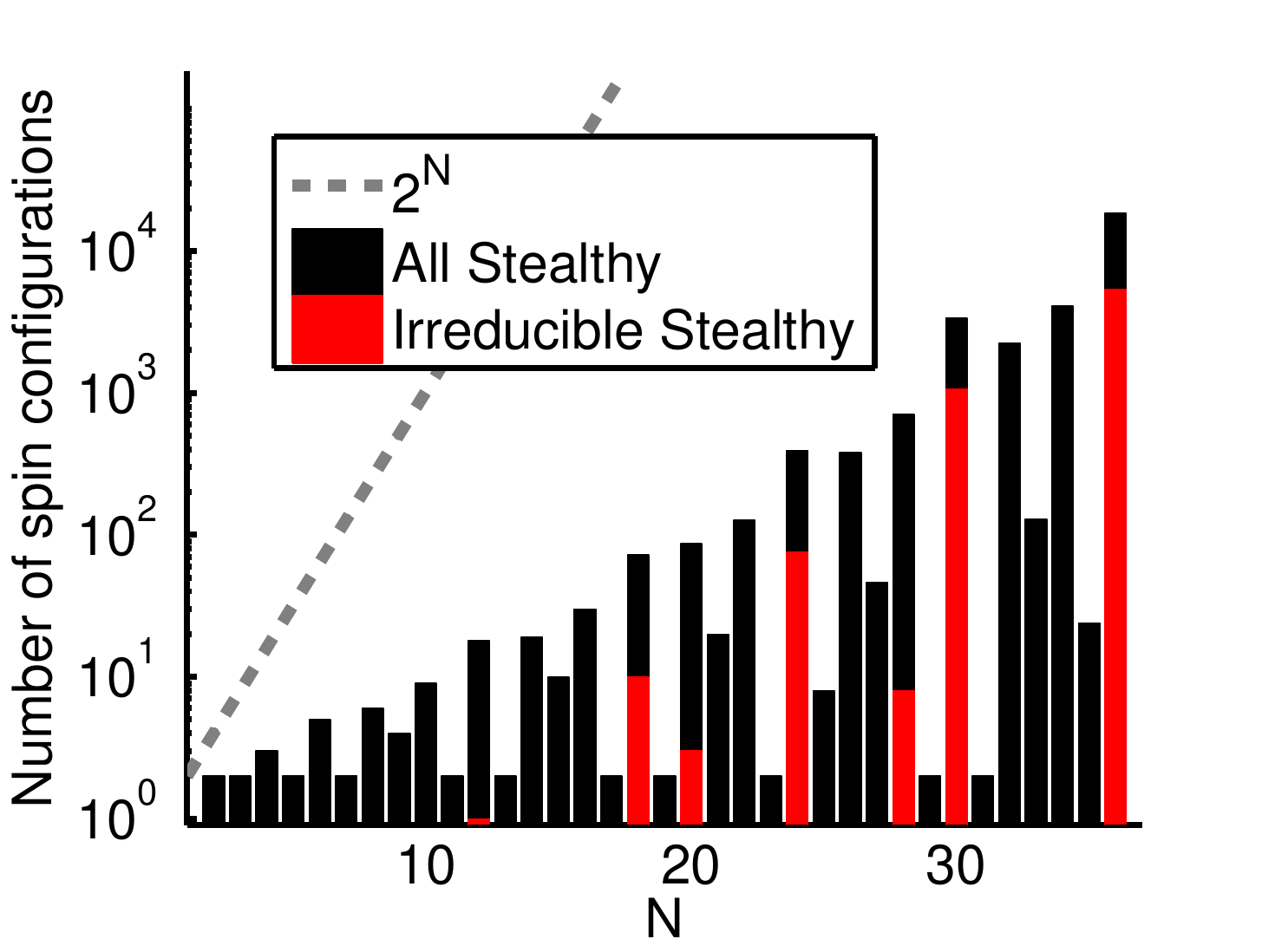} & \includegraphics[width=0.4\textwidth]{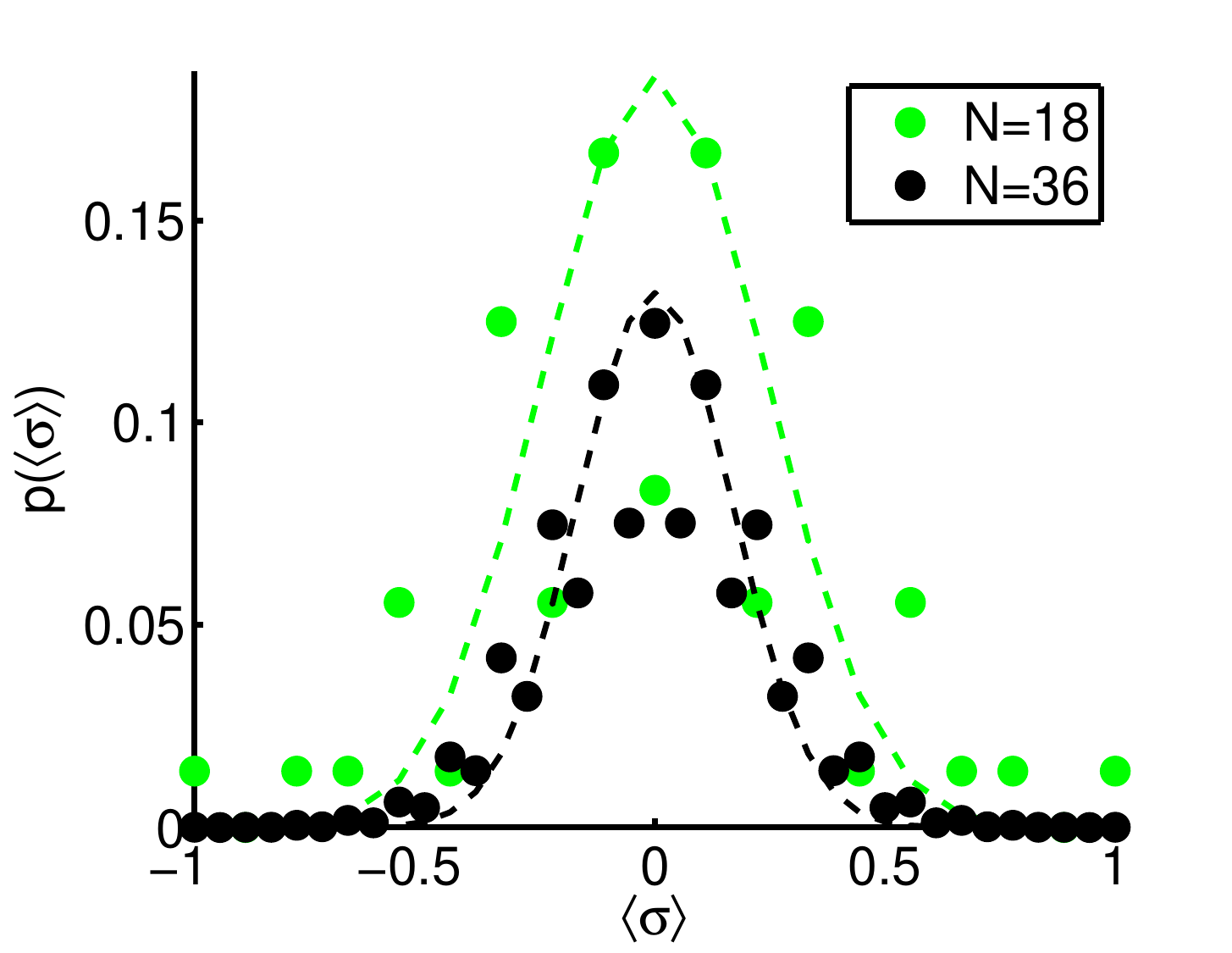} \\
(a) & (b)
\end{tabular}
\end{center}
\caption{(a) The number of stealthy hyperuniform spin configurations for spin configurations of periodic unit cell sizes between $N=2$ and $36$. In gray, for comparison, is the line $2^N$. In black are all of the stealthy hyperuniform configurations. In red are the irreducible stealthy configurations, which cannot be expressed in unit cells smaller than $N$. (b) The distribution of magnetization for irreducible spin configurations. For comparison, the symmetric binomial distributions about $\langle\sigma\rangle=0$ are shown as dashed lines.} \label{fig:stealthyenumeration}
\end{figure*}

The enumerated stealthy spin configurations can be expressed in terms of spin configurations of fixed size and magnetization, two defining properties of generic spin patterns. Of primary interest is how the number of stealthy spins grows with system size. Figure~\ref{fig:stealthyenumeration}(a) displays the abundance of stealthy hyperuniform spin configurations relative to all possible $2^N$ configurations. The general trend for stealthy spin configurations of size $N$ is exponential growth $a^N$ in the number of configurations at a rate of $a\approx 1.4 < 2$. Spin configurations with prime $N$ have only the two trivial stealthy configurations of all up spins and all down spins.  Moreover, most stealthy spin configurations of size $N\leq 36$ are reducible to fundamental unit cells of size less than $N$. As discussed in the Appendix, this property comes from the fact that all stealthy spin configurations of $N=pq$, where $p$ and $q$ are prime, are reducible to smaller bases. The values of $N$ that do not follow this pattern admit irreducible stealthy configurations. They are shown in red in Fig.~\ref{fig:stealthyenumeration}(a) and are labeled ``irreducible.''

The probability distribution of magnetization for configurations of stealthy spins of size $N$, shown in Fig.~\ref{fig:stealthyenumeration}(b), is symmetric and binomial-like, though with some key differences. For finite system sizes, depending on $N$, a spin configuration distribution is peaked at either $\langle\sigma\rangle=0$ or $\langle\sigma\rangle=\pm 1/N$. This behavior is determined by the prime factorization of $N$ and the abundance of reducible configurations of size $N$. For comparison, Fig.~\ref{fig:stealthyenumeration}(b) shows the magnetization distribution of all spin configurations on chains of length $N$, which exactly follows a binomial distribution. It is interesting to note that the distributions of magnetization of the stealthy configurations tend to oscillate about the binomial profile. This seems to be a finite size effect that has to do with the fact that more stealthy configurations with an even number of up spins exist than with an odd number.

\subsubsection{Order and disorder in stealthy spin configurations}

\begin{figure*}[t]
\begin{center}
\begin{tabular}{cc}
\includegraphics[width=0.45\textwidth]{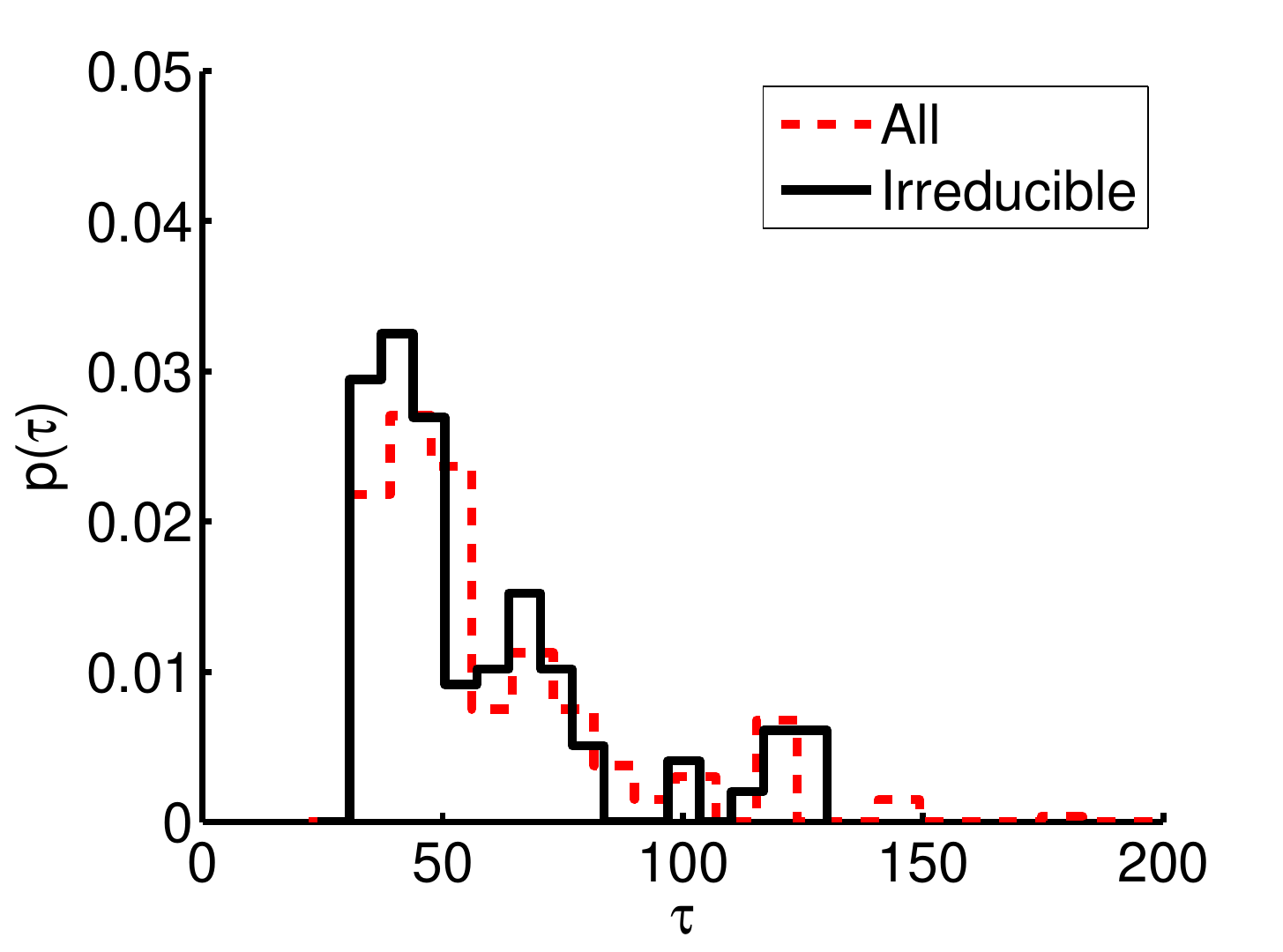} & \includegraphics[width=0.45\textwidth]{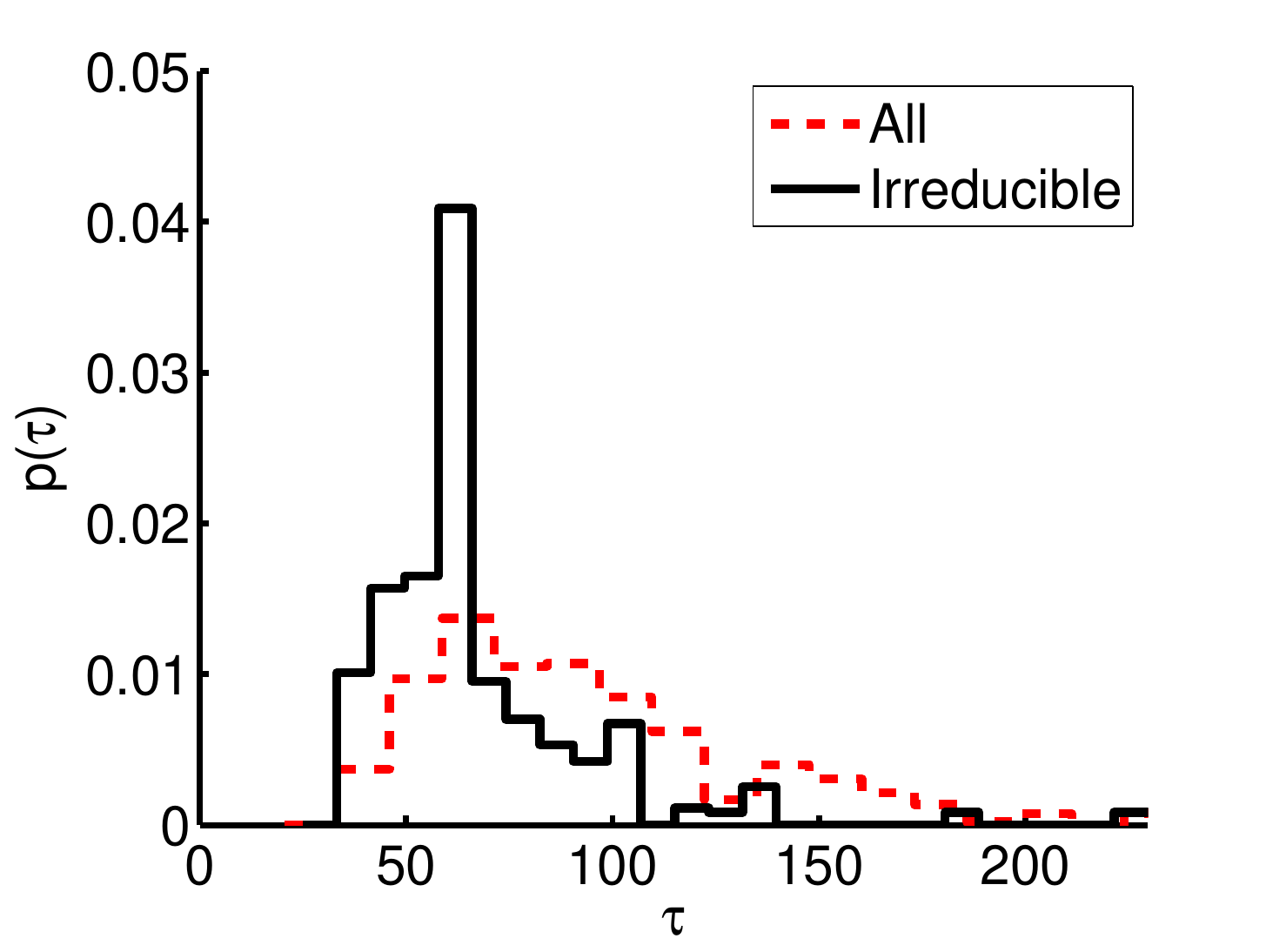} \\
(a) & (b)
\end{tabular}
\end{center}
\caption{Histograms representing the probability density of the order metric $\tau$ for stealthy hyperuniform spin chains of size (a) $N=30$ and (b) $N=36$ with zero magnetization. Displayed in dashed red lines are histograms of the $\tau$ distribution among \emph{all} stealthy patterns of the specified sizes, including the reducible configurations which can be expressed in terms of smaller unit cells. A few extremely ordered configurations with $\tau > 500$ are not shown. Displayed in solid black lines are histograms of the $\tau$ distribution for irreducible stealthy configurations. Note that there are 2306 size $N=36$ stealthy configuration, 872 of which are irreducible, and that there are 315 size $N=30$ stealthy configurations, 297 of which are irreducible.} \label{fig:taudist}
\end{figure*}

Our enumeration reveals a range of disorder for hyperuniform stealthy spin systems. The distributions of disorder for stealthy spin configurations, according to the order metric defined in Eq. (\ref{eq:tau}), are displayed in Fig.~\ref{fig:taudist}. Figure~\ref{fig:taudist}(a) shows the distribution of the order metric $\tau$ among irreducible only and all stealthy spin configurations of size $N=30$ and $\langle\sigma\rangle=0$. The distribution of $\tau$ tends to spread out as the system size increases and is positively skewed. Figure~\ref{fig:taudist}(b) shows the distribution of $\tau$ for $N=36$ and $\langle\sigma\rangle=0$ stealthy configurations and highlights the importance of irreducibility in stealthy spin systems. For size $N=36$, the irreducible configuration are the most disordered of all the stealthy configurations.

\begin{figure}[t]
\begin{center}
\includegraphics[width=0.4\textwidth]{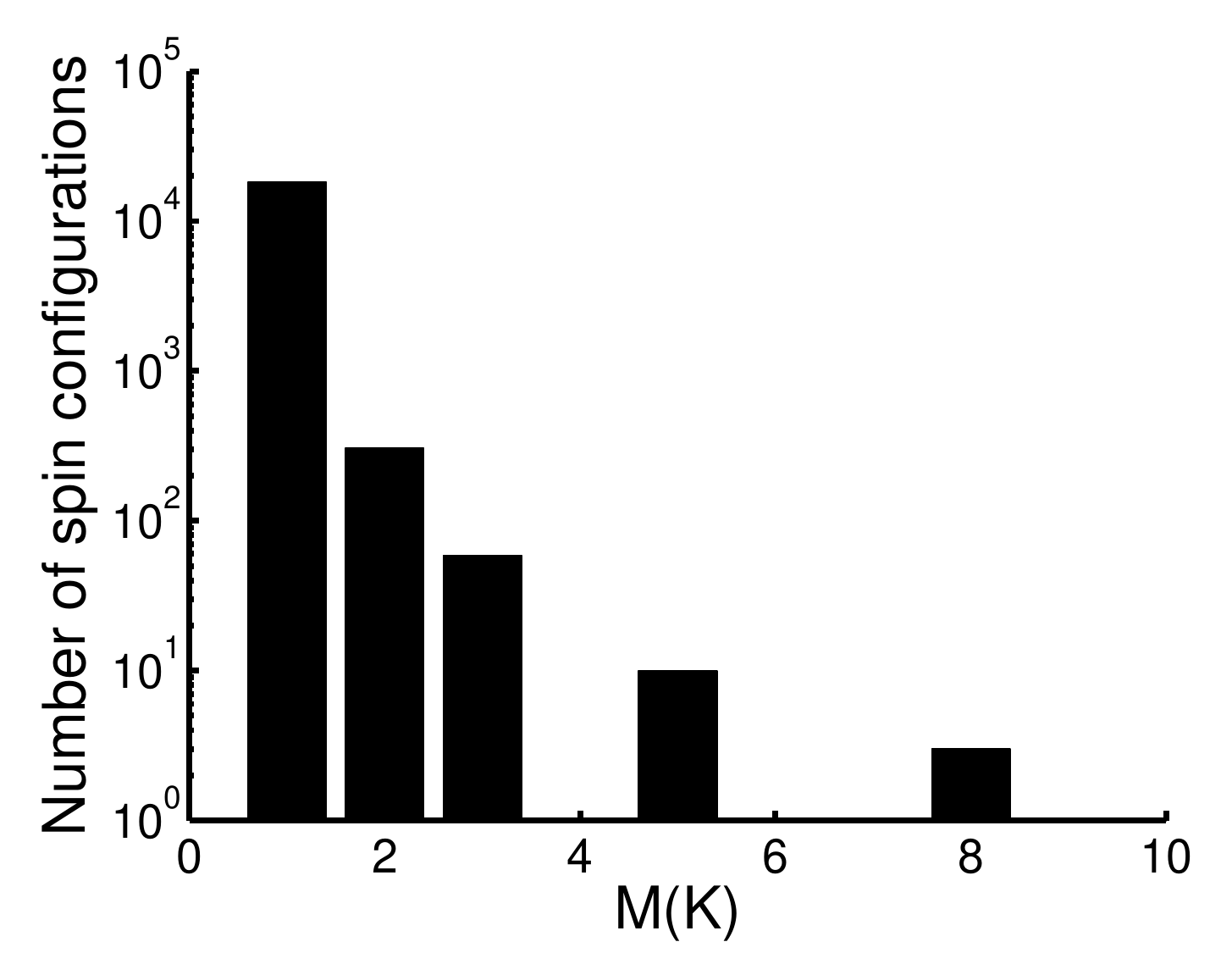}%
\end{center}
\caption{The distribution of $M(K)$ [number of wavevectors for which $S(k)=0$ for $k \leq K$] for all enumerated stealthy configurations of unit cell size $N=36$.} \label{fig:exclusionradius}
\end{figure}

The nature of disorder in our 1D spin systems is intimately linked to their stealthiness. We measure the degree of stealthiness of a material by the number of independent wavevectors for which $S(k)=0$ for $k\leq K$, which for 1D periodic spins chains on the integer lattice is $M(K)=NK/2\pi$, proportional to exclusion zone radius $K$. The distribution of $M(K)$ for all stealthy spins of size $N=36$ is shown in Fig.~\ref{fig:exclusionradius}. We see that the distribution of $M(k)$ is skewed towards the smaller values of $K$. In general, larger $M(K)$ correspond to configurations with more Bragg peaks in the range of wavenumbers $0 < k < 2\pi$ and hence larger $\tau$ values. This is particularly true for the irreducible stealthy spin configurations, which only have exclusion radii with $M(K)\leq 2$ and are the most disordered stealthy configurations. This correlation between stealthiness and order in spin chains, along with similar evidence of order in stealthy continuous particle systems with large exclusion zone radii,~\cite{stealthy2} suggests a strong positive relationship between the absence of magnetic and radiation scattering for large wavelengths and ordering. Disordered stealthy spin-chains require a relatively small exclusion zone radius and a narrower range of wavevectors for which scattering does not occur.

\subsection{Groundstateability}

By application of the inverse statistical mechanics method developed in Ref.~\onlinecite{designer1}, we have been able to show that irreducible stealthy hyperuniform spin chains can be stabilized as ground states of the long-ranged (\textit{i.e.}, relative to the chain length), spin-spin interaction defined in Eq.~(\ref{eq:H}). The ``groundstateability'' of these irreducible configurations were classified into class I, II, and III, according to whether those configurations could be uniquely realized as ground states or not, as detailed in Sec.~\ref{sec:methods}. 

\begin{figure}[t]
\begin{center}
\begin{tabular}{c}
\includegraphics[width=0.4\textwidth]{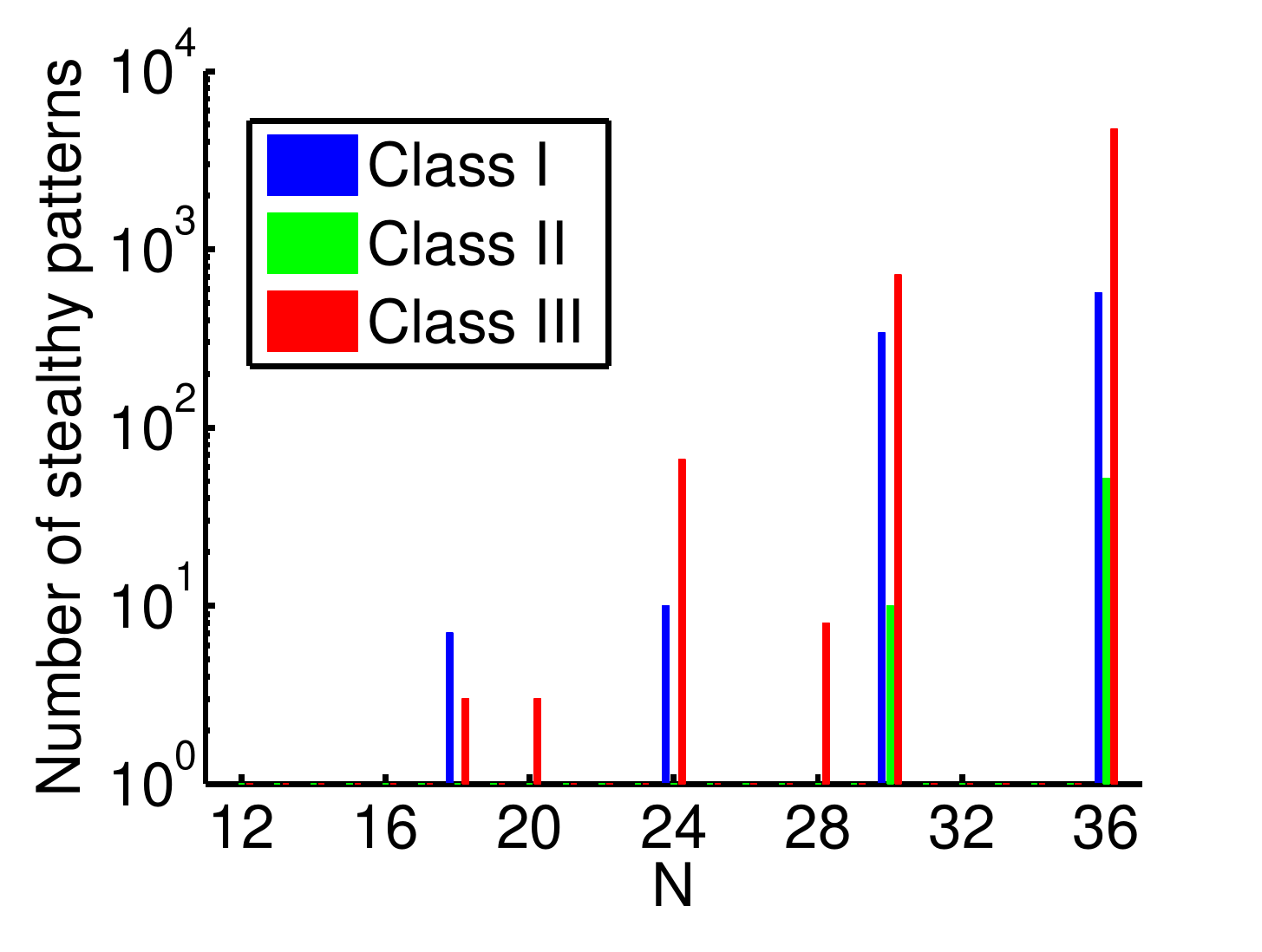} \\
(a) \\
\includegraphics[width=0.4\textwidth]{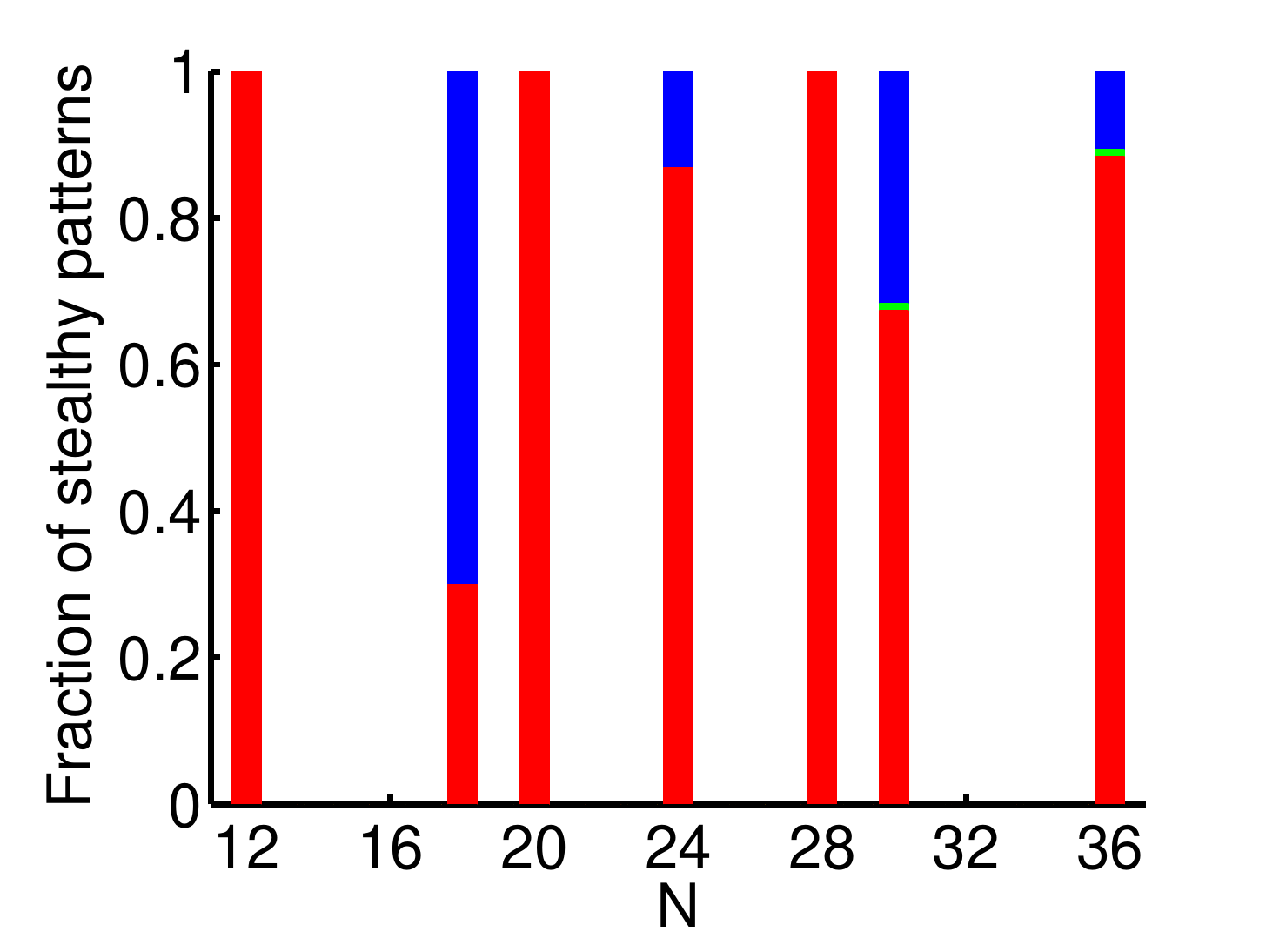} \\
(b)
\end{tabular}
\end{center}
\caption{(a) The number of irreducible stealthy 1D spin configurations in classes I, II, and III for sizes $N=12,\ldots,36$ plotted on a log-scale. The first size with class II configurations is $N=30$. $N=12$ has one class III configuration. Configurations equivalent by a spin-inversion are not counted. (b) The relative amounts of these configurations in classes I, II, and III. Most of the sizes studied have no irreducible stealthy configurations and so are not depicted. Only 10 out of 1068 irreducible stealthy configurations with period $N=30$ and 52 out of 5362 irreducible stealthy configurations with period $N=36$ are class II.} \label{fig:stealthyclasses}
\end{figure}

Figure~\ref{fig:stealthyclasses} displays the ground-state classification of irreducible stealthy spin configurations. As demonstrated in Fig.~\ref{fig:taudist}, these configurations are the most disordered of the stealthy spin configurations at these sizes. For the given stabilizing potential and limited spin chain lengths considered here, a majority of stealthy configurations are class III. Nonetheless, a surprisingly large fraction of stealthy configurations at $N=18,24,30,36$ can be uniquely stabilized. Among the stabilized configurations, most are class I. Class II stealthy configurations are rare, but begin to appear for the larger system sizes. It has been demonstrated that, for 1D spin chains on the integer lattice, as $N$ increases the number of $S_2$ degeneracies increases.~\cite{designer2} This trend holds also for stealthy spin chains and seems to indicate that more class II configurations occur for larger $N$. This trend of increasing degeneracy among stealthy hyperuniform configurations would agree with the recent results for the particle analogs.~\cite{2003_torquato}

We should note that the number of class I and II configurations might not be quite accurate due to the heuristic nature of the zero-temperature competitor-based method implemented.~\cite{designer1} Class III configurations cannot be incorrectly classified, as discussed in Sec.~\ref{sec:methods}. However, true class III configurations can be incorrectly classified as class I or II if the competitor space is not adequately sampled by the simulated annealing procedure. Incorrect classification is more likely for large fundamental unit cells $N$ and large potential potential cutoffs $R_C$, where the solution space becomes large and full of local minima. Nonetheless, we took care to verify the class I and II configurations found through the inverse method. After obtaining a solution, we ran 10 -- 50 of iterations of simulated annealing optimizations to confirm the target configuration as the ground state. Each simulated annealing run started with different initial conditions and involved 10,000 -- 30,000 Monte Carlo sweeps.

\begin{figure}[t]
\begin{center}
\includegraphics[width=0.45\textwidth]{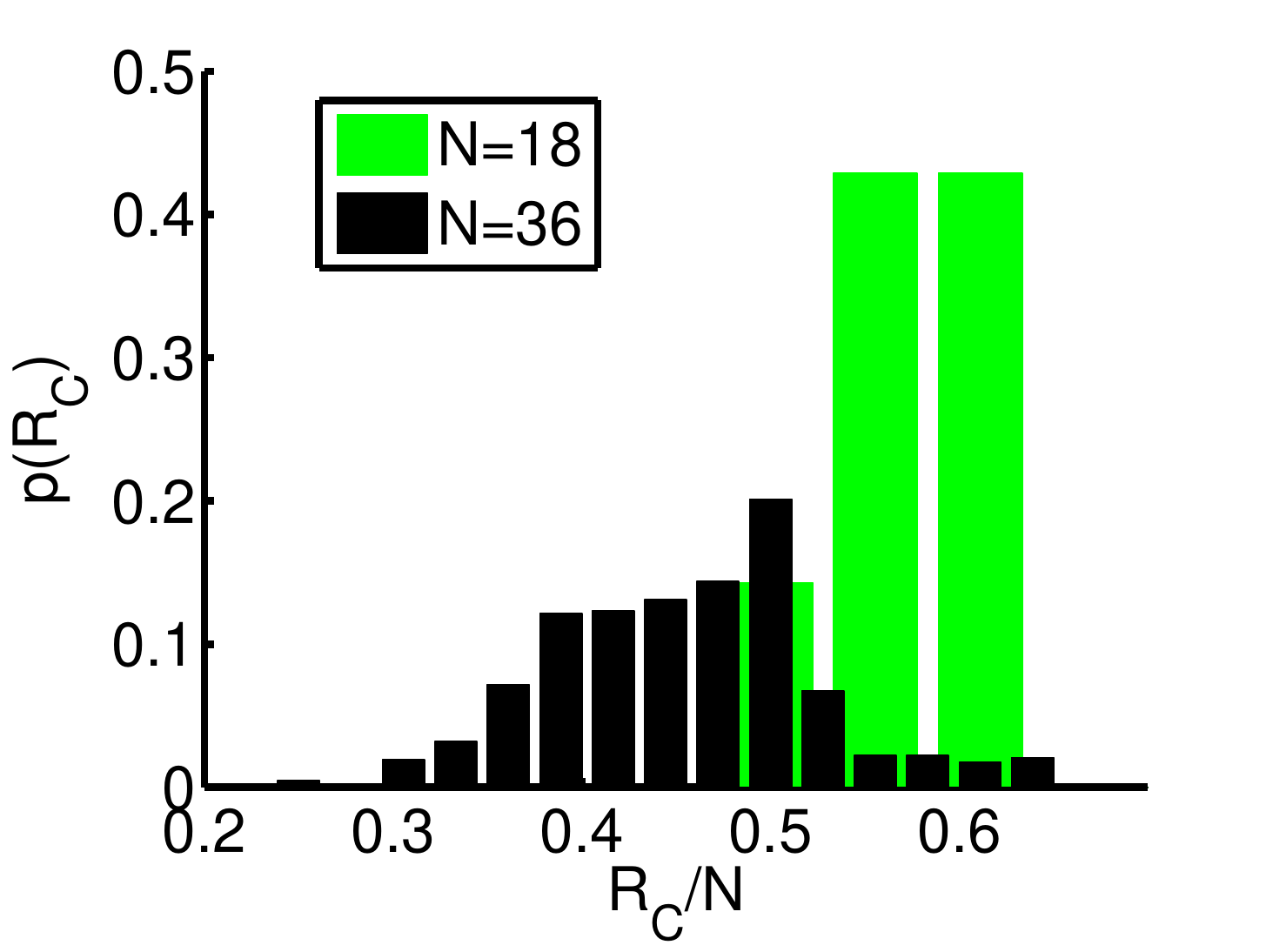} 
\end{center}
\caption{The probability distributions of relative potential cutoffs $R_C/N$, where $R_C$ is an integer between 1 and $N$, for $N=18$ and $N=36$ periodic 1D irreducible stealthy spins systems on the integer lattice. Larger systems tend to have a greater spread in relative potential cutoff centered around lower values. For $N=18$ and $N=24$ (not shown), all potential cutoffs were at least half of the period length $R_C\geq N/2$. While $N=30$ (not shown) and $N=36$ contained many configurations which could be represented as ground states of potentials of length $R_C < N/2$.} \label{fig:cutoffs}
\end{figure}

The $J(R)$ potentials engineered by the inverse method are the shortest possible stabilizing interactions. Interestingly, these interactions are long-ranged, relative to the chain length $N$, and display a surprising range of cutoff distances, with $9 \leq R_C \leq 23$ for size $N=36$ stealthy spin chains. Figure~\ref{fig:cutoffs} shows the distributions of potential cutoffs $R_C$ for Class I and Class II irreducible stealthy spin configurations of size $N=18$ and $N=36$. An important characteristic of these distributions is that the variance in the range of the potential increases with $N$. This is promising for the future engineering of disordered stealthy spin configurations. It would seem plausible for short-range ($R_C/N \ll 1$) interactions to produce disorder, stealthiness, and hyperuniformity.

\begin{figure}[t]
\begin{center}
\begin{tabular}{|c|}
\hline (a) \\
\includegraphics[width=0.45\textwidth]{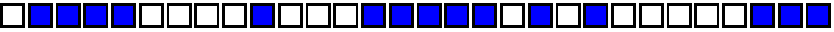} \\
\includegraphics[width=0.45\textwidth]{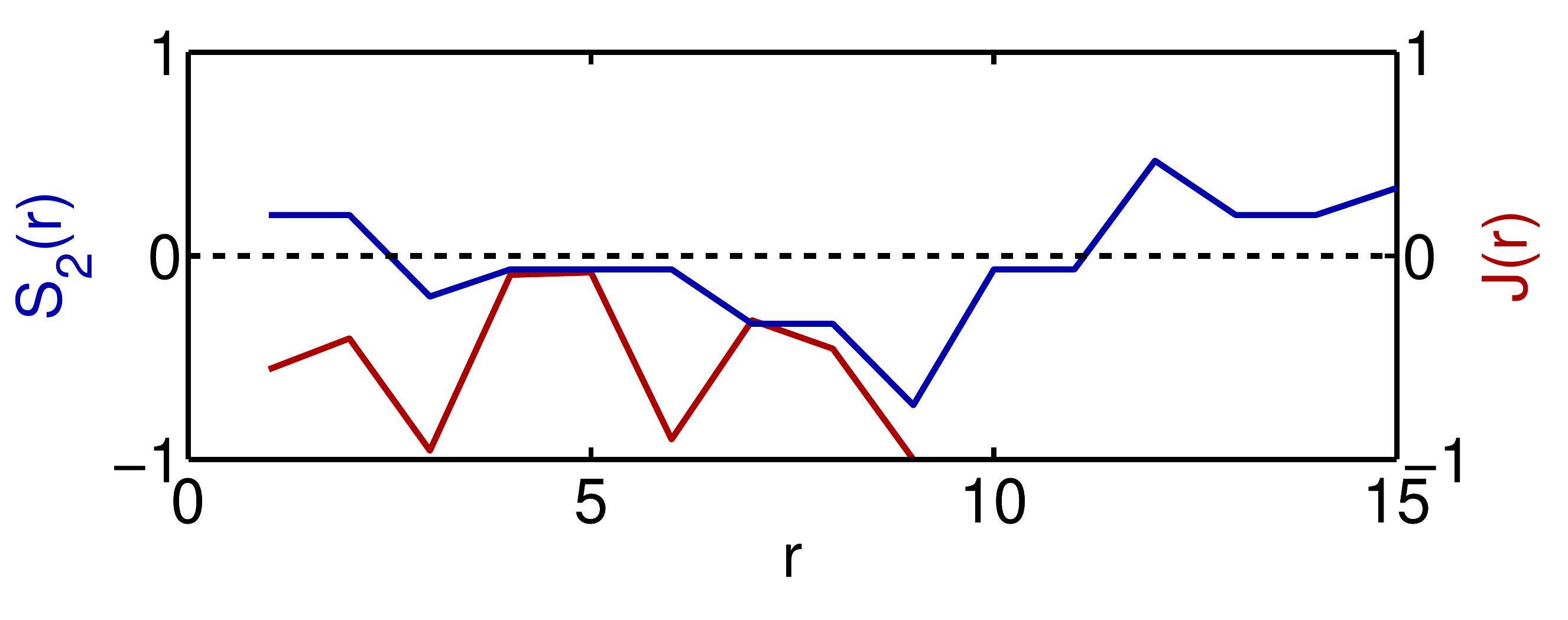} \\
\hline (b) \\
\includegraphics[width=0.45\textwidth]{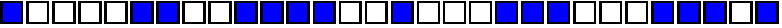} \\
\includegraphics[width=0.45\textwidth]{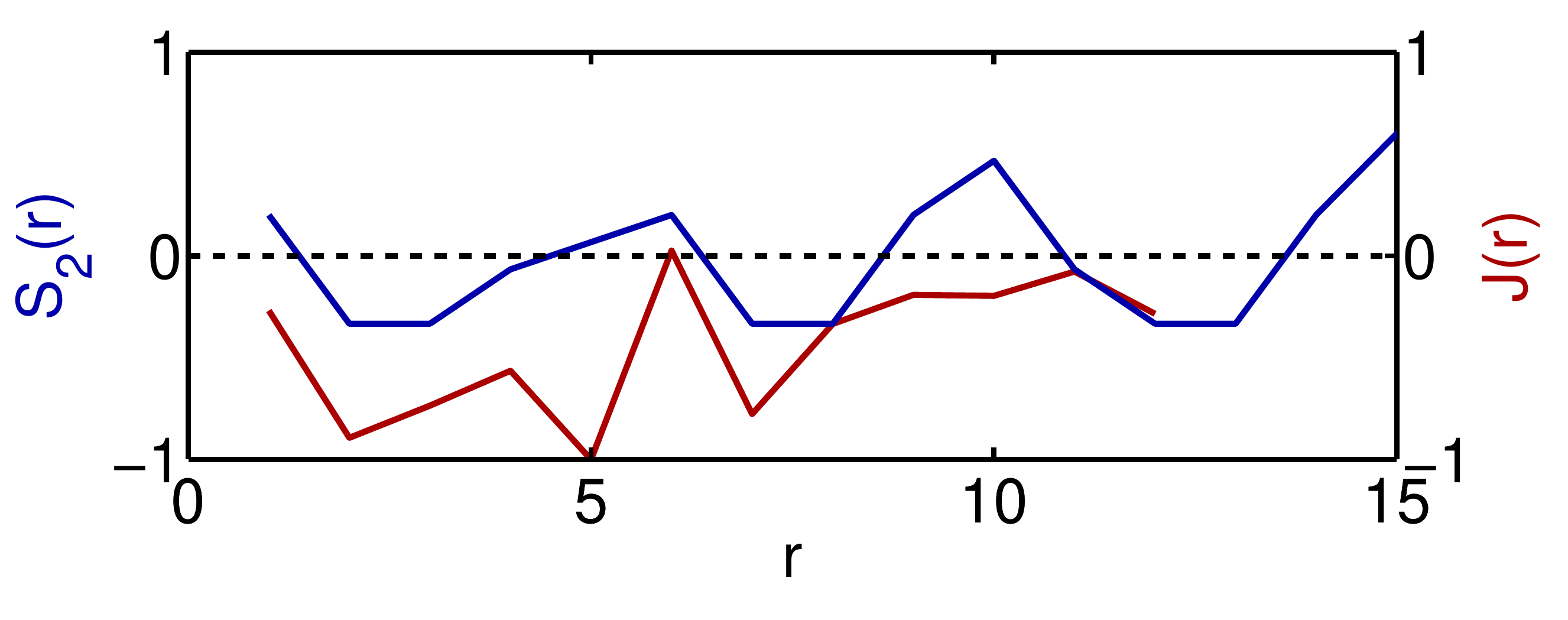} \\
\hline (c) \\
\includegraphics[width=0.45\textwidth]{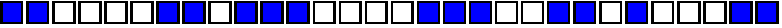} \\
\includegraphics[width=0.45\textwidth]{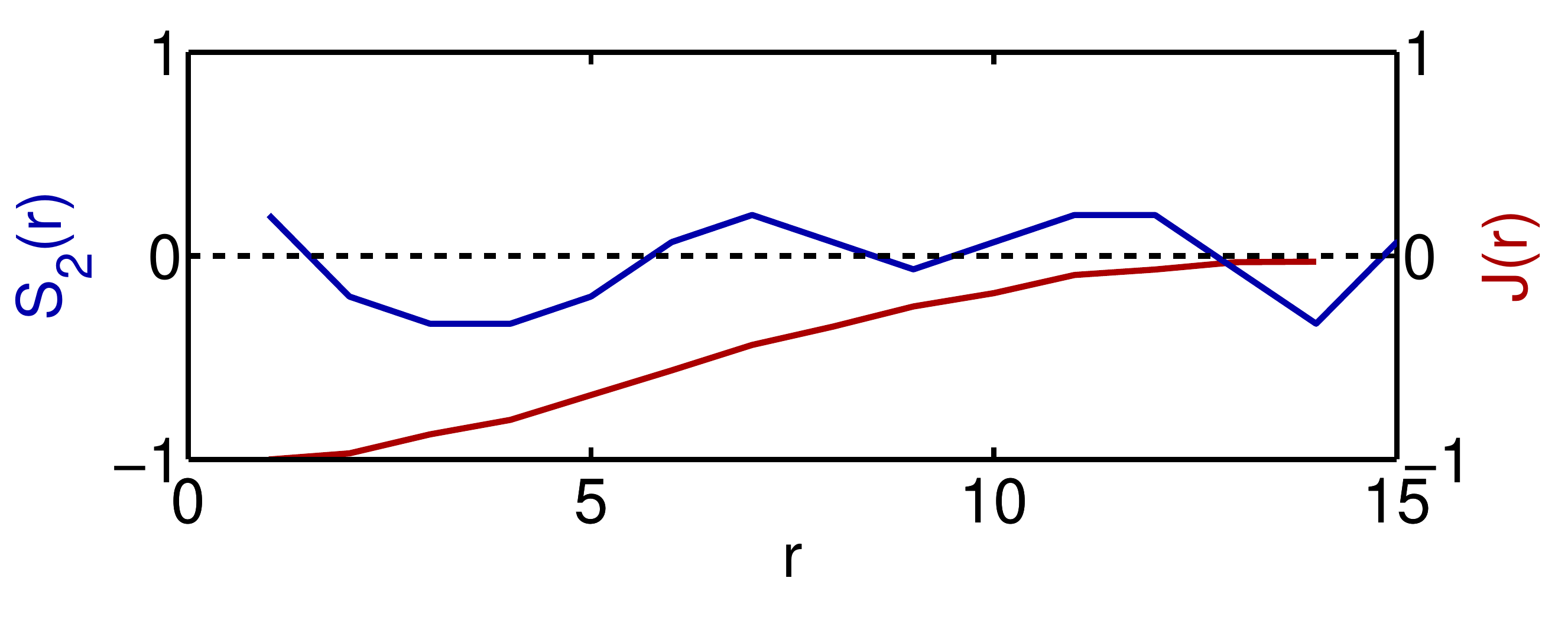} \\
\hline
\end{tabular}
\end{center}
\caption{Three representative stealthy class I spin configurations of size 30 with $\langle\sigma\rangle=0$. Depicted are their unique spin-spin correlations $S_2(R)$ in blue and their stabilizing potentials $J(R)$ in red with potential cutoffs (a)$R_C=9$, (b) $R_C=12$, and (c) $R_C=14$. Note that the $J(R)$ are not unique and that other stabilizing interactions could exist, though not with cutoffs less than the given $R_C$.} \label{fig:classI_potentials}
\end{figure}

There is a large variety of interactions that can stabilize stealthy hyperuniform spin chains as ground states. To showcase this variety in size $N=36$ spin chains, we pick three potentials with $R_C$ from different points ($R_C=9,12,14$) in the distribution depicted in Fig.~\ref{fig:cutoffs}. Figure~\ref{fig:classI_potentials} shows the three stabilizing potentials, along with the class I stealthy spin configurations they generate and the spin-spin correlation of the configurations. Figure~\ref{fig:classI_potentials}(a) shows a $\langle\sigma\rangle=0$ configuration generated by a short ($R_C=9$) potential. Both the spin-spin correlation and the potential $J(R)$ vary greatly on short length scales. Figure~\ref{fig:classI_potentials}(b) shows a longer ($R_C=12$) potential, which, along with the spin-spin correlation, changes more smoothly as a function of distance. The longest potential displayed ($R_C=14$) in Figure~\ref{fig:classI_potentials}(c) shows the smoothest stabilizing potential and spin-correlation of the three configurations shown. These three class I spin configurations and their corresponding stabilizing potentials are representative.

\begin{figure*}[t]
\begin{center}
\begin{tabular}{cc}
\includegraphics[width=0.48\textwidth]{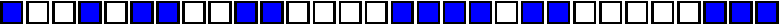} & \includegraphics[width=0.48\textwidth]{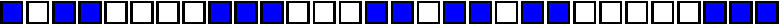} \\ \\
\includegraphics[width=0.48\textwidth]{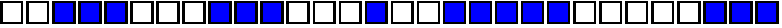} & \includegraphics[width=0.48\textwidth]{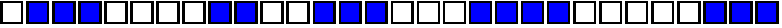} \\ \\
\includegraphics[width=0.48\textwidth]{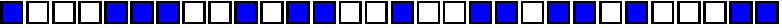} & \includegraphics[width=0.48\textwidth]{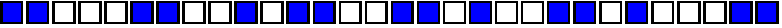} \\ \\
\includegraphics[width=0.48\textwidth]{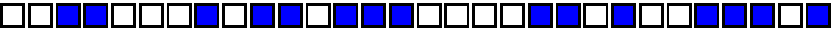} & \includegraphics[width=0.48\textwidth]{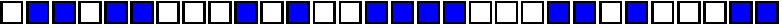} \\ \\
\includegraphics[width=0.48\textwidth]{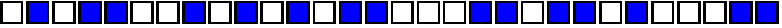} & \includegraphics[width=0.48\textwidth]{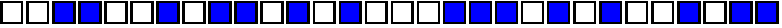}
\end{tabular}
\end{center}
\caption{All class II stealthy spin configurations of size $30$. Five pairs of $S_2(R)$-degenerate configurations are grouped horizontally.} \label{fig:size30_classII}
\end{figure*}

Our characterization of class II configurations is more limited. We have only managed to observe a few dozen such configurations for size $30$ and $36$ spin chains. All size 30 class II stealthy configurations are shown in Fig.~\ref{fig:size30_classII}. For size 30 and 36 stealthy hyperuniform spin chains, all class II configurations came in $S_2$-degenerate pairs, rather than in larger degenerate groups. It seems likely that for larger $N$, degeneracy should increase given the large combinatorial increase in the number of possible configurations and ways to construct stealthy, hyperuniform patterns. 

\section{Conclusions} \label{sec:conclusions}

In this work, we have shown that disordered stealthy hyperuniform spin configurations can be realized as either unique or degenerate ground states of radial long-ranged (relative to the chain length) spin-spin interactions.
Stealthy hyperuniform states are unique in that they are transparent to radiation for a range of wavenumbers around the origin, implying anomalously suppressed magnetization fluctuations at long wavelengths. The discovered exotic disordered spin ground states, distinctly different from spin glasses, are the spin analogs of disordered stealthy hyperuniform many-particle ground states~\cite{stealthydisorder,torquato_arxiv} that have been shown to be endowed with novel photonic properties.~\cite{quantumcascadelaser,2009_torquato,isotropicbandgaps}
Thus, stealthy hyperuniform spin systems offer potentially exciting new avenues for future research, as we will elaborate below.

It is useful to summarize how we came to ascertain that such disordered spin ground states exist.
First, we performed an exhaustive enumeration over the entire set of spin configurations that exist on periodic 1D integer lattices containing $N=2,3,\ldots,36$ sites in order to identify and characterize \textit{all} stealthy hyperuniform spin chains in this system size range.
In doing so, we found that the number of stealthy hyperuniform spin configurations grows exponentially with system size (\textit{i.e.}, $\sim 1.4^N$), implying that the fraction of these unique spin configurations goes to zero in the thermodynamic limit.
Furthermore, the distributions of magnetization and order in these stealthy hyperuniform spin configurations show strong deviations from the binomial distributions of these quantities that are characteristic of the set of all enumerated spin configurations.

To study the ``groundstateability'' of disordered stealthy hyperuniform spin configurations, we employed recently developed inverse statistical mechanics techniques~\cite{designer1,designer2} in conjunction with a class of Hamiltonians that allow for radial pairwise spin-spin interactions that extend well beyond nearest-neighbor lattice sites.
Although many of these spin configurations cannot be stabilized as unique (Class I) or $S_2(R)$-degenerate (Class II) solutions within this set of allowed spin-spin interaction types, we did identify a significant number of Class I and II disordered stealthy hyperuniform spin chains, in particular for the largest lattices considered herein.
Interestingly, the spin-spin interaction potentials that were able to spontaneously produce these disordered spin chains display a wide radial extent, spanning from $\approx$ 30\% to $\approx$ 60\% of the length of the entire underlying integer lattice.
Although these long-range interactions (with respect to the size of the lattice) tend to be attractive over their defined radial extent, there still exists a great variety in the shapes and relative magnitudes of these spin-spin interaction potentials. 
As such, these findings are promising indicators that such interactions---and therefore such exotic spin configurations---can in fact be realized experimentally.

Our fundamental understanding of disordered stealthy hyperuniform spin configurations and the spin-spin interaction potentials required to stabilize them is currently in its infancy.
Hence the implications of the existence of disordered stealthy hyperuniform spin ground states established herein provide fertile ground for future research directions.
We are interested in determining whether the excited states and bulk physical properties of these systems are singularly remarkable and characterizing these potential novel states of matter.
The ground-state classification of these systems in the thermodynamic limit still remains an outstanding open question as well as the effects of thermal and/or quantum mechanical fluctuations on their stealthy and hyperuniformity properties.
Such knowledge would be invaluable in realizing the rational design of these exotic spin systems and exploring their potential technological applications.

\begin{acknowledgments}
R. D. and R. C. were supported by the U.S. Department of Energy, Office of Basic Energy Sciences, under DE-FG02-05ER46201.
\end{acknowledgments}

\appendix*
\section{Reducibility of stealthy hyperuniform spin chains} \label{ap:trivialstealthiness}

Most stealthy hyperuniform spin chains on the integer lattice are represented in a larger than necessary fundamental unit cell. These so called \emph{reducible} configurations can be defined with a smaller fundamental unit cell, under which they are not stealthy. Consider, for instance, the antiferromagnetic configuration depicted at the top of Fig.~\ref{fig:confs_order}. This configuration is not stealthy, i.e. does not have $S(k)=0$ at the smallest positive wavenumber $k=2\pi/N$, when represented in a periodic fundamental unit cell of size $N=2$.

The unusual trend in the number of irreducible stealthy configurations as a function of fundamental unit cell size $N$, shown in Fig.~\ref{fig:stealthyenumeration}(a), suggests that the reducibility of stealthy configurations is related to the prime factorization of $N$. While we could not easily identify the general trend, we were able to characterize the reducibility of a particularly simple factorization of $N$.

\emph{Conjecture:} A spin chain that is stealthy hyperuniform on the integer lattice with fundamental unit cell size $N=pq$, where $p$ and $q$ are prime, is always reducible, i.e. can be equivalently represented by a smaller unit cell of size $N'<N$ in which it does not appear stealthy.

\emph{Discussion:} We use the notation $\sigma = (\sigma_1, \sigma_2, \ldots, \sigma_N)$ to signify spin configurations. At size $N$, one can easily construct two crystalline configurations $\sigma^{(p)}$ and $\sigma^{(q)}$ that cause the collective density variables, and hence the structure factor, to vanish when $k=2\pi/N$. First, we will show this for the $\sigma^{(p)}$ configuration
\begin{align}
\sigma_j^{(p)} &= \begin{cases} +1 & j=p,2p,\ldots,qp \\ -1 & \mbox{otherwise} \end{cases}.
\end{align}
Its collective density variables at the smallest positive wavenumber can be expressed as
\begin{align}
\rho^{(p)}(2\pi/N) &= \sum_{j=1}^N \sigma_j^{(p)}e^{i(2\pi j/N)} \\
&= \sum_{n=1}^{q} e^{i(2\pi np/N)} - \sum_{j\neq np}^{N} e^{i(2\pi j/N)} \\
&= \rho^{(p)}_+(2\pi/N) - \rho^{(p)}_-(2\pi/N)
\end{align}
We see that the second term can be expressed in terms of the first term
\begin{align}
\rho^{(p)}_-(2\pi/N) &= \sum_{n=1}^{q} e^{i(2\pi (np+1)/N)} + \cdots + \sum_{n=1}^{q} e^{i(2\pi (np+(p-1))/N)} \\
&= \sum_{m=1}^{p-1} e^{i(2\pi m/N)} \rho^{(p)}_+(2\pi/N)
\end{align}
where we used the periodicity of the chain.

Now we can see that if $\rho^{(p)}_+$ vanishes, so does $\rho^{(p)}_-$. As it turns out, the $\rho^{(p)}_+$ term does indeed evaluate to zero:
\begin{align}
\rho^{(p)}_+(2\pi/N) &= \sum_{n=1}^{q} e^{i(2\pi np/N)} \\
&= \frac{1}{1-e^{-i(2\pi p/N)}}\left( 1 - e^{i(2\pi pq/N)}\right) \\
&= \frac{1}{1-e^{-i(2\pi p/N)}}\left( 1 - e^{i(2\pi)}\right) \\
&= 0.
\end{align}
Therefore, $\rho^{(p)}(2\pi/N) = 0$ and $\sigma^{(p)}$ is stealthy. The same logic applies to $\sigma^{(q)}$ defined as
\begin{align}
\sigma_j^{(q)} &= \begin{cases} +1 & j=q,2q,\ldots,pq \\ -1 & \mbox{otherwise} \end{cases}
\end{align}
so we can conclude that it is also stealthy.

Both $\sigma^{(p)}$ and $\sigma^{(q)}$ are reducible. One might think that a new irreducible stealthy configuration $\sigma'$ could be generated by superposing $\sigma^{(p)}$ and $\sigma^{(q)}$, taking $\sigma'_j = +1$ when $\sigma^{(p)}_j = +1$ or $\sigma^{(q)}_j = +1$.  Based on the same reason presented in Sec.~VI of Ref.~\onlinecite{torquato_arxiv}, the superposed configuration is stealthy if the sequences $\{p,2p,\ldots,qp \}$ and $\{q,2q,\ldots,pq \}$ do not overlap. However, they do overlap exactly once at $pq$. Since the collective density variables are linear in their terms, this leaves $\rho'(2\pi/N)$ with one term in its summation that does not cancel. Therefore, $\rho'(2\pi/N) \neq 0$ and the overlapped configuration is not stealthy.

Moreover, our enumeration results suggest that $\sigma^{(p)}$ and $\sigma^{(q)}$ are the only stealthy configurations that exist for $N=pq$ (modulo spin inversion and translation). Both are trivial configurations that are reducible to fundamental unit cells of size $N'=p$ and $q$ respectively. In the reduced $N'$ cell, these chains have only a single up spin which leads to a term that does not cancel in the density $\rho(2\pi/N')$. Therefore, the reduced configurations are not stealthy.

Therefore, all stealthy hyperuniform spin chains of size $N=pq$ are reducible.

\bibliography{literature}

\begin{thebibliography}{54}%
\makeatletter
\providecommand \@ifxundefined [1]{%
 \@ifx{#1\undefined}
}%
\providecommand \@ifnum [1]{%
 \ifnum #1\expandafter \@firstoftwo
 \else \expandafter \@secondoftwo
 \fi
}%
\providecommand \@ifx [1]{%
 \ifx #1\expandafter \@firstoftwo
 \else \expandafter \@secondoftwo
 \fi
}%
\providecommand \natexlab [1]{#1}%
\providecommand \enquote  [1]{``#1''}%
\providecommand \bibnamefont  [1]{#1}%
\providecommand \bibfnamefont [1]{#1}%
\providecommand \citenamefont [1]{#1}%
\providecommand \href@noop [0]{\@secondoftwo}%
\providecommand \href [0]{\begingroup \@sanitize@url \@href}%
\providecommand \@href[1]{\@@startlink{#1}\@@href}%
\providecommand \@@href[1]{\endgroup#1\@@endlink}%
\providecommand \@sanitize@url [0]{\catcode `\\12\catcode `\$12\catcode
  `\&12\catcode `\#12\catcode `\^12\catcode `\_12\catcode `\%12\relax}%
\providecommand \@@startlink[1]{}%
\providecommand \@@endlink[0]{}%
\providecommand \url  [0]{\begingroup\@sanitize@url \@url }%
\providecommand \@url [1]{\endgroup\@href {#1}{\urlprefix }}%
\providecommand \urlprefix  [0]{URL }%
\providecommand \Eprint [0]{\href }%
\providecommand \doibase [0]{http://dx.doi.org/}%
\providecommand \selectlanguage [0]{\@gobble}%
\providecommand \bibinfo  [0]{\@secondoftwo}%
\providecommand \bibfield  [0]{\@secondoftwo}%
\providecommand \translation [1]{[#1]}%
\providecommand \BibitemOpen [0]{}%
\providecommand \bibitemStop [0]{}%
\providecommand \bibitemNoStop [0]{.\EOS\space}%
\providecommand \EOS [0]{\spacefactor3000\relax}%
\providecommand \BibitemShut  [1]{\csname bibitem#1\endcsname}%
\let\auto@bib@innerbib\@empty
\bibitem [{\citenamefont {Torquato}(2009)}]{inverseopt}%
  \BibitemOpen
  \bibfield  {author} {\bibinfo {author} {\bibfnamefont {S.}~\bibnamefont
  {Torquato}},\ }\href {\doibase 10.1039/B814211B} {\bibfield  {journal}
  {\bibinfo  {journal} {Soft Matter}\ }\textbf {\bibinfo {volume} {5}},\
  \bibinfo {pages} {1157} (\bibinfo {year} {2009})}\BibitemShut {NoStop}%
\bibitem [{\citenamefont {Ising}(1925)}]{ising}%
  \BibitemOpen
  \bibfield  {author} {\bibinfo {author} {\bibfnamefont {E.}~\bibnamefont
  {Ising}},\ }\href {\doibase 10.1007/BF02980577} {\bibfield  {journal}
  {\bibinfo  {journal} {Z. Phys.}\ }\textbf {\bibinfo {volume} {31}},\ \bibinfo
  {pages} {253} (\bibinfo {year} {1925})}\BibitemShut {NoStop}%
\bibitem [{\citenamefont {{F. Y. Wu}}(1982)}]{potts}%
  \BibitemOpen
  \bibfield  {author} {\bibinfo {author} {\bibnamefont {{F. Y. Wu}}},\ }\href
  {\doibase 10.1103/RevModPhys.54.235} {\bibfield  {journal} {\bibinfo
  {journal} {Rev. Mod. Phys.}\ }\textbf {\bibinfo {volume} {54}},\ \bibinfo
  {pages} {235} (\bibinfo {year} {1982})}\BibitemShut {NoStop}%
\bibitem [{\citenamefont {Onsager}(1944)}]{onsager}%
  \BibitemOpen
  \bibfield  {author} {\bibinfo {author} {\bibfnamefont {L.}~\bibnamefont
  {Onsager}},\ }\href {\doibase 10.1103/PhysRev.65.117} {\bibfield  {journal}
  {\bibinfo  {journal} {Phys. Rev.}\ }\textbf {\bibinfo {volume} {65}},\
  \bibinfo {pages} {117} (\bibinfo {year} {1944})}\BibitemShut {NoStop}%
\bibitem [{\citenamefont {{C. N. Yang}}(1952)}]{yang}%
  \BibitemOpen
  \bibfield  {author} {\bibinfo {author} {\bibnamefont {{C. N. Yang}}},\ }\href
  {\doibase 10.1103/PhysRev.85.808} {\bibfield  {journal} {\bibinfo  {journal}
  {Phys. Rev.}\ }\textbf {\bibinfo {volume} {85}},\ \bibinfo {pages} {808}
  (\bibinfo {year} {1952})}\BibitemShut {NoStop}%
\bibitem [{\citenamefont {{M. C. Rechtsman}}\ \emph {et~al.}(2005)\citenamefont
  {{M. C. Rechtsman}}, \citenamefont {{F. H. Stillinger}},\ and\ \citenamefont
  {Torquato}}]{torquato_honeycomb}%
  \BibitemOpen
  \bibfield  {author} {\bibinfo {author} {\bibnamefont {{M. C. Rechtsman}}},
  \bibinfo {author} {\bibnamefont {{F. H. Stillinger}}}, \ and\ \bibinfo
  {author} {\bibfnamefont {S.}~\bibnamefont {Torquato}},\ }\href {\doibase
  10.1103/PhysRevLett.95.228301} {\bibfield  {journal} {\bibinfo  {journal}
  {Phys. Rev. Lett.}\ }\textbf {\bibinfo {volume} {95}},\ \bibinfo {pages}
  {228301} (\bibinfo {year} {2005})}\BibitemShut {NoStop}%
\bibitem [{\citenamefont {Rechtsman}\ \emph {et~al.}(2006)\citenamefont
  {Rechtsman}, \citenamefont {Stillinger},\ and\ \citenamefont
  {Torquato}}]{torquato_triangularsquarehoneycomb}%
  \BibitemOpen
  \bibfield  {author} {\bibinfo {author} {\bibfnamefont {M.}~\bibnamefont
  {Rechtsman}}, \bibinfo {author} {\bibfnamefont {F.}~\bibnamefont
  {Stillinger}}, \ and\ \bibinfo {author} {\bibfnamefont {S.}~\bibnamefont
  {Torquato}},\ }\href {\doibase 10.1103/PhysRevE.73.011406} {\bibfield
  {journal} {\bibinfo  {journal} {Phys. Rev. E}\ }\textbf {\bibinfo {volume}
  {73}},\ \bibinfo {pages} {011406} (\bibinfo {year} {2006})}\BibitemShut
  {NoStop}%
\bibitem [{\citenamefont {Cohn}\ and\ \citenamefont {Kumar}(2009)}]{cohnkumar}%
  \BibitemOpen
  \bibfield  {author} {\bibinfo {author} {\bibfnamefont {H.}~\bibnamefont
  {Cohn}}\ and\ \bibinfo {author} {\bibfnamefont {A.}~\bibnamefont {Kumar}},\
  }\href {\doibase 10.1073/pnas.0901636106} {\bibfield  {journal} {\bibinfo
  {journal} {Proc. Natl. Acad. Sci. USA}\ }\textbf {\bibinfo {volume} {160}},\
  \bibinfo {pages} {9570} (\bibinfo {year} {2009})}\BibitemShut {NoStop}%
\bibitem [{\citenamefont {{\'{E}. Marcotte}}\ \emph {et~al.}(2011)\citenamefont
  {{\'{E}. Marcotte}}, \citenamefont {{F. H. Stillinger}},\ and\ \citenamefont
  {Torquato}}]{torquato_monotonicsquarehoneycomb}%
  \BibitemOpen
  \bibfield  {author} {\bibinfo {author} {\bibnamefont {{\'{E}. Marcotte}}},
  \bibinfo {author} {\bibnamefont {{F. H. Stillinger}}}, \ and\ \bibinfo
  {author} {\bibfnamefont {S.}~\bibnamefont {Torquato}},\ }\href {\doibase
  10.1039/C0SM01205J} {\bibfield  {journal} {\bibinfo  {journal} {Soft Matter}\
  }\textbf {\bibinfo {volume} {7}},\ \bibinfo {pages} {2332} (\bibinfo {year}
  {2011})}\BibitemShut {NoStop}%
\bibitem [{\citenamefont {{M. C. Rechtsman}}\ \emph
  {et~al.}(2007{\natexlab{a}})\citenamefont {{M. C. Rechtsman}}, \citenamefont
  {{F. H. Stillinger}},\ and\ \citenamefont {Torquato}}]{torquato_diamond}%
  \BibitemOpen
  \bibfield  {author} {\bibinfo {author} {\bibnamefont {{M. C. Rechtsman}}},
  \bibinfo {author} {\bibnamefont {{F. H. Stillinger}}}, \ and\ \bibinfo
  {author} {\bibfnamefont {S.}~\bibnamefont {Torquato}},\ }\href {\doibase
  10.1103/PhysRevE.75.031403} {\bibfield  {journal} {\bibinfo  {journal} {Phys.
  Rev. E}\ }\textbf {\bibinfo {volume} {75}},\ \bibinfo {pages} {031403}
  (\bibinfo {year} {2007}{\natexlab{a}})}\BibitemShut {NoStop}%
\bibitem [{\citenamefont {{C. L. M{\"u}ller}}\ and\ \citenamefont {{I. F.
  Sbalzarini}}(2012)}]{muller}%
  \BibitemOpen
  \bibfield  {author} {\bibinfo {author} {\bibnamefont {{C. L. M{\"u}ller}}}\
  and\ \bibinfo {author} {\bibnamefont {{I. F. Sbalzarini}}},\ }\href {\doibase
  10.1162/EVCO_a_00086} {\bibfield  {journal} {\bibinfo  {journal} {Evol.
  Comput.}\ }\textbf {\bibinfo {volume} {20}},\ \bibinfo {pages} {543}
  (\bibinfo {year} {2012})}\BibitemShut {NoStop}%
\bibitem [{\citenamefont {{A. F. Hannon}}\ \emph {et~al.}(2014)\citenamefont
  {{A. F. Hannon}}, \citenamefont {Ding}, \citenamefont {Bai}, \citenamefont
  {{C. A. Ross}},\ and\ \citenamefont {Alexander{-}Katz}}]{hannon}%
  \BibitemOpen
  \bibfield  {author} {\bibinfo {author} {\bibnamefont {{A. F. Hannon}}},
  \bibinfo {author} {\bibfnamefont {Y.}~\bibnamefont {Ding}}, \bibinfo {author}
  {\bibfnamefont {W.}~\bibnamefont {Bai}}, \bibinfo {author} {\bibnamefont {{C.
  A. Ross}}}, \ and\ \bibinfo {author} {\bibfnamefont {A.}~\bibnamefont
  {Alexander{-}Katz}},\ }\href {\doibase 10.1021/nl404067s} {\bibfield
  {journal} {\bibinfo  {journal} {Nano Lett.}\ }\textbf {\bibinfo {volume}
  {14}},\ \bibinfo {pages} {318} (\bibinfo {year} {2014})}\BibitemShut
  {NoStop}%
\bibitem [{\citenamefont {Zhang}\ \emph {et~al.}(2013)\citenamefont {Zhang},
  \citenamefont {{F. H. Stillinger}},\ and\ \citenamefont {Torquato}}]{zhang}%
  \BibitemOpen
  \bibfield  {author} {\bibinfo {author} {\bibfnamefont {G.}~\bibnamefont
  {Zhang}}, \bibinfo {author} {\bibnamefont {{F. H. Stillinger}}}, \ and\
  \bibinfo {author} {\bibfnamefont {S.}~\bibnamefont {Torquato}},\ }\href
  {\doibase 10.1103/PhysRevE.88.042309} {\bibfield  {journal} {\bibinfo
  {journal} {Phys. Rev. E}\ }\textbf {\bibinfo {volume} {88}},\ \bibinfo
  {pages} {042309} (\bibinfo {year} {2013})}\BibitemShut {NoStop}%
\bibitem [{\citenamefont {Jain}\ \emph
  {et~al.}(2014{\natexlab{a}})\citenamefont {Jain}, \citenamefont {{J. R.
  Errington}},\ and\ \citenamefont {{T. M. Truskett}}}]{jain1}%
  \BibitemOpen
  \bibfield  {author} {\bibinfo {author} {\bibfnamefont {A.}~\bibnamefont
  {Jain}}, \bibinfo {author} {\bibnamefont {{J. R. Errington}}}, \ and\
  \bibinfo {author} {\bibnamefont {{T. M. Truskett}}},\ }\href {\doibase
  10.1103/PhysRevX.4.031049} {\bibfield  {journal} {\bibinfo  {journal} {Phys.
  Rev. X}\ }\textbf {\bibinfo {volume} {4}},\ \bibinfo {pages} {031049}
  (\bibinfo {year} {2014}{\natexlab{a}})}\BibitemShut {NoStop}%
\bibitem [{\citenamefont {Jain}\ \emph
  {et~al.}(2014{\natexlab{b}})\citenamefont {Jain}, \citenamefont {{J. A.
  Bollinger}},\ and\ \citenamefont {{T. M. Truskett}}}]{jain2}%
  \BibitemOpen
  \bibfield  {author} {\bibinfo {author} {\bibfnamefont {A.}~\bibnamefont
  {Jain}}, \bibinfo {author} {\bibnamefont {{J. A. Bollinger}}}, \ and\
  \bibinfo {author} {\bibnamefont {{T. M. Truskett}}},\ }\href {\doibase
  10.1002/aic.14491} {\bibfield  {journal} {\bibinfo  {journal} {AiChE
  Journal}\ }\textbf {\bibinfo {volume} {60}},\ \bibinfo {pages} {2732}
  (\bibinfo {year} {2014}{\natexlab{b}})}\BibitemShut {NoStop}%
\bibitem [{\citenamefont {{M. C. Rechtsman}}\ \emph
  {et~al.}(2007{\natexlab{b}})\citenamefont {{M. C. Rechtsman}}, \citenamefont
  {{F. H. Stillinger}},\ and\ \citenamefont
  {Torquato}}]{negativethermalexpansion}%
  \BibitemOpen
  \bibfield  {author} {\bibinfo {author} {\bibnamefont {{M. C. Rechtsman}}},
  \bibinfo {author} {\bibnamefont {{F. H. Stillinger}}}, \ and\ \bibinfo
  {author} {\bibfnamefont {S.}~\bibnamefont {Torquato}},\ }\href
  {http://arxiv.org/pdf/0807.3559.pdf} {\bibfield  {journal} {\bibinfo
  {journal} {J. Phys. Chem. A}\ }\textbf {\bibinfo {volume} {111}},\ \bibinfo
  {pages} {12816} (\bibinfo {year} {2007}{\natexlab{b}})}\BibitemShut {NoStop}%
\bibitem [{\citenamefont {{M. C. Rechtsman}}\ \emph {et~al.}(2008)\citenamefont
  {{M. C. Rechtsman}}, \citenamefont {{F. H. Stillinger}},\ and\ \citenamefont
  {Torquato}}]{negativepoissonratio}%
  \BibitemOpen
  \bibfield  {author} {\bibinfo {author} {\bibnamefont {{M. C. Rechtsman}}},
  \bibinfo {author} {\bibnamefont {{F. H. Stillinger}}}, \ and\ \bibinfo
  {author} {\bibfnamefont {S.}~\bibnamefont {Torquato}},\ }\href {\doibase
  10.1103/PhysRevLett.101.085501} {\bibfield  {journal} {\bibinfo  {journal}
  {Phys. Rev. Lett.}\ }\textbf {\bibinfo {volume} {101}},\ \bibinfo {pages}
  {085501} (\bibinfo {year} {2008})}\BibitemShut {NoStop}%
\bibitem [{\citenamefont {Franceschetti}\ and\ \citenamefont
  {Zunger}(1999)}]{zunger1}%
  \BibitemOpen
  \bibfield  {author} {\bibinfo {author} {\bibfnamefont {A.}~\bibnamefont
  {Franceschetti}}\ and\ \bibinfo {author} {\bibfnamefont {A.}~\bibnamefont
  {Zunger}},\ }\href {\doibase 10.1038/46995} {\bibfield  {journal} {\bibinfo
  {journal} {Nature}\ }\textbf {\bibinfo {volume} {402}},\ \bibinfo {pages}
  {60} (\bibinfo {year} {1999})}\BibitemShut {NoStop}%
\bibitem [{\citenamefont {{G. L. W. Hart}}\ \emph {et~al.}(2005)\citenamefont
  {{G. L. W. Hart}}, \citenamefont {{V. Blum}}, \citenamefont {{M. J.
  Walorski}},\ and\ \citenamefont {Zunger}}]{zunger2}%
  \BibitemOpen
  \bibfield  {author} {\bibinfo {author} {\bibnamefont {{G. L. W. Hart}}},
  \bibinfo {author} {\bibnamefont {{V. Blum}}}, \bibinfo {author} {\bibnamefont
  {{M. J. Walorski}}}, \ and\ \bibinfo {author} {\bibfnamefont
  {A.}~\bibnamefont {Zunger}},\ }\href {\doibase 10.1038/nmat1374} {\bibfield
  {journal} {\bibinfo  {journal} {Nat. Mater.}\ }\textbf {\bibinfo {volume}
  {4}},\ \bibinfo {pages} {391} (\bibinfo {year} {2005})}\BibitemShut {NoStop}%
\bibitem [{\citenamefont {Blum}\ \emph {et~al.}(2005)\citenamefont {Blum},
  \citenamefont {{G. L. W. Hart}}, \citenamefont {{M. J. Walorski}},\ and\
  \citenamefont {Zunger}}]{zunger3}%
  \BibitemOpen
  \bibfield  {author} {\bibinfo {author} {\bibfnamefont {V.}~\bibnamefont
  {Blum}}, \bibinfo {author} {\bibnamefont {{G. L. W. Hart}}}, \bibinfo
  {author} {\bibnamefont {{M. J. Walorski}}}, \ and\ \bibinfo {author}
  {\bibfnamefont {A.}~\bibnamefont {Zunger}},\ }\href {\doibase
  10.1103/PhysRevB.72.165113} {\bibfield  {journal} {\bibinfo  {journal} {Phys.
  Rev. B}\ }\textbf {\bibinfo {volume} {72}},\ \bibinfo {pages} {165113}
  (\bibinfo {year} {2005})}\BibitemShut {NoStop}%
\bibitem [{\citenamefont {{R. A. DiStasio Jr.}}\ \emph
  {et~al.}(2013)\citenamefont {{R. A. DiStasio Jr.}}, \citenamefont {Marcotte},
  \citenamefont {Car}, \citenamefont {{F. H. Stillinger}},\ and\ \citenamefont
  {Torquato}}]{designer1}%
  \BibitemOpen
  \bibfield  {author} {\bibinfo {author} {\bibnamefont {{R. A. DiStasio Jr.}}},
  \bibinfo {author} {\bibfnamefont {E.}~\bibnamefont {Marcotte}}, \bibinfo
  {author} {\bibfnamefont {R.}~\bibnamefont {Car}}, \bibinfo {author}
  {\bibnamefont {{F. H. Stillinger}}}, \ and\ \bibinfo {author} {\bibfnamefont
  {S.}~\bibnamefont {Torquato}},\ }\href {\doibase 10.1103/PhysRevB.88.134104}
  {\bibfield  {journal} {\bibinfo  {journal} {Phys. Rev. B}\ }\textbf {\bibinfo
  {volume} {88}},\ \bibinfo {pages} {134104} (\bibinfo {year}
  {2013})}\BibitemShut {NoStop}%
\bibitem [{\citenamefont {{\'{E}. Marcotte}}\ \emph {et~al.}(2013)\citenamefont
  {{\'{E}. Marcotte}}, \citenamefont {{R. A. DiStasio Jr.}}, \citenamefont {{F.
  H. Stillinger}},\ and\ \citenamefont {Torquato}}]{designer2}%
  \BibitemOpen
  \bibfield  {author} {\bibinfo {author} {\bibnamefont {{\'{E}. Marcotte}}},
  \bibinfo {author} {\bibnamefont {{R. A. DiStasio Jr.}}}, \bibinfo {author}
  {\bibnamefont {{F. H. Stillinger}}}, \ and\ \bibinfo {author} {\bibfnamefont
  {S.}~\bibnamefont {Torquato}},\ }\href {\doibase 10.1103/PhysRevB.88.184432}
  {\bibfield  {journal} {\bibinfo  {journal} {Phys. Rev. B}\ }\textbf {\bibinfo
  {volume} {88}},\ \bibinfo {pages} {184432} (\bibinfo {year}
  {2013})}\BibitemShut {NoStop}%
\bibitem [{\citenamefont {{F. A. Ma'Mari}}\ \emph {et~al.}(2015)\citenamefont
  {{F. A. Ma'Mari}}, \citenamefont {Moorsom}, \citenamefont {Teobaldi},
  \citenamefont {Deacon}, \citenamefont {Prokscha}, \citenamefont {Luetkens},
  \citenamefont {Lee}, \citenamefont {{G. E. Sterbinsky}}, \citenamefont {{D.
  A. Arena}}, \citenamefont {{D. A. MacLaren}}, \citenamefont {Flokstra},
  \citenamefont {Ali}, \citenamefont {{M. C. Wheeler}}, \citenamefont
  {Burnell}, \citenamefont {{B. J. Hickey}},\ and\ \citenamefont
  {Cespedes}}]{stoner}%
  \BibitemOpen
  \bibfield  {author} {\bibinfo {author} {\bibnamefont {{F. A. Ma'Mari}}},
  \bibinfo {author} {\bibfnamefont {T.}~\bibnamefont {Moorsom}}, \bibinfo
  {author} {\bibfnamefont {G.}~\bibnamefont {Teobaldi}}, \bibinfo {author}
  {\bibfnamefont {W.}~\bibnamefont {Deacon}}, \bibinfo {author} {\bibfnamefont
  {T.}~\bibnamefont {Prokscha}}, \bibinfo {author} {\bibfnamefont
  {H.}~\bibnamefont {Luetkens}}, \bibinfo {author} {\bibfnamefont
  {S.}~\bibnamefont {Lee}}, \bibinfo {author} {\bibnamefont {{G. E.
  Sterbinsky}}}, \bibinfo {author} {\bibnamefont {{D. A. Arena}}}, \bibinfo
  {author} {\bibnamefont {{D. A. MacLaren}}}, \bibinfo {author} {\bibfnamefont
  {M.}~\bibnamefont {Flokstra}}, \bibinfo {author} {\bibfnamefont
  {M.}~\bibnamefont {Ali}}, \bibinfo {author} {\bibnamefont {{M. C. Wheeler}}},
  \bibinfo {author} {\bibfnamefont {G.}~\bibnamefont {Burnell}}, \bibinfo
  {author} {\bibnamefont {{B. J. Hickey}}}, \ and\ \bibinfo {author}
  {\bibfnamefont {O.}~\bibnamefont {Cespedes}},\ }\href {\doibase
  10.1038/nature14621} {\bibfield  {journal} {\bibinfo  {journal} {Nature}\
  }\textbf {\bibinfo {volume} {524}},\ \bibinfo {pages} {69} (\bibinfo {year}
  {2015})}\BibitemShut {NoStop}%
\bibitem [{\citenamefont {Torquato}\ and\ \citenamefont {{F. H.
  Stillinger}}(2003)}]{2003_torquato}%
  \BibitemOpen
  \bibfield  {author} {\bibinfo {author} {\bibfnamefont {S.}~\bibnamefont
  {Torquato}}\ and\ \bibinfo {author} {\bibnamefont {{F. H. Stillinger}}},\
  }\href {\doibase 10.1103/PhysRevE.68.041113} {\bibfield  {journal} {\bibinfo
  {journal} {Phys. Rev. E}\ }\textbf {\bibinfo {volume} {68}},\ \bibinfo
  {pages} {041113} (\bibinfo {year} {2003})}\BibitemShut {NoStop}%
\bibitem [{\citenamefont {{C. E. Zachary}}\ and\ \citenamefont
  {Torquato}(2009)}]{hyperuniformpointpatterns}%
  \BibitemOpen
  \bibfield  {author} {\bibinfo {author} {\bibnamefont {{C. E. Zachary}}}\ and\
  \bibinfo {author} {\bibfnamefont {S.}~\bibnamefont {Torquato}},\ }\href
  {\doibase 10.1088/1742-5468/2009/12/P12015} {\bibfield  {journal} {\bibinfo
  {journal} {J. Stat. Mech. Theor. Exp.}\ }\textbf {\bibinfo {volume} {12}},\
  \bibinfo {pages} {15} (\bibinfo {year} {2009})}\BibitemShut {NoStop}%
\bibitem [{\citenamefont {Gabrielli}\ \emph {et~al.}(2003)\citenamefont
  {Gabrielli}, \citenamefont {Jancovici}, \citenamefont {Joyce}, \citenamefont
  {{J. L. Lebowitz}}, \citenamefont {Pietronero},\ and\ \citenamefont {{F. S.
  Labini}}}]{statisticalphysicscosmic}%
  \BibitemOpen
  \bibfield  {author} {\bibinfo {author} {\bibfnamefont {A.}~\bibnamefont
  {Gabrielli}}, \bibinfo {author} {\bibfnamefont {B.}~\bibnamefont
  {Jancovici}}, \bibinfo {author} {\bibfnamefont {M.}~\bibnamefont {Joyce}},
  \bibinfo {author} {\bibnamefont {{J. L. Lebowitz}}}, \bibinfo {author}
  {\bibfnamefont {L.}~\bibnamefont {Pietronero}}, \ and\ \bibinfo {author}
  {\bibnamefont {{F. S. Labini}}},\ }\href {\doibase
  10.1103/PhysRevD.67.043506} {\bibfield  {journal} {\bibinfo  {journal} {Phys.
  Rev. D}\ }\textbf {\bibinfo {volume} {67}},\ \bibinfo {pages} {043506}
  (\bibinfo {year} {2003})}\BibitemShut {NoStop}%
\bibitem [{\citenamefont {{P. J. E. Peebles}}(1993)}]{peebles}%
  \BibitemOpen
  \bibfield  {author} {\bibinfo {author} {\bibnamefont {{P. J. E. Peebles}}},\
  }\href {http://www.sciencedirect.com/science/book/9780123870322} {\emph
  {\bibinfo {title} {Principles of Physical Cosmology}}},\ \bibinfo {edition}
  {1st}\ ed.\ (\bibinfo  {publisher} {Princeton University Press, Princeton,
  NJ},\ \bibinfo {year} {1993})\BibitemShut {NoStop}%
\bibitem [{\citenamefont {Jiao}\ \emph {et~al.}(2014)\citenamefont {Jiao},
  \citenamefont {Lau}, \citenamefont {Hatzikiriou}, \citenamefont
  {Meyer{-}Hermann}, \citenamefont {{J. C. Corbo}},\ and\ \citenamefont
  {Torquato}}]{avianphotoreceptors}%
  \BibitemOpen
  \bibfield  {author} {\bibinfo {author} {\bibfnamefont {Y.}~\bibnamefont
  {Jiao}}, \bibinfo {author} {\bibfnamefont {T.}~\bibnamefont {Lau}}, \bibinfo
  {author} {\bibfnamefont {H.}~\bibnamefont {Hatzikiriou}}, \bibinfo {author}
  {\bibfnamefont {M.}~\bibnamefont {Meyer{-}Hermann}}, \bibinfo {author}
  {\bibnamefont {{J. C. Corbo}}}, \ and\ \bibinfo {author} {\bibfnamefont
  {S.}~\bibnamefont {Torquato}},\ }\href {\doibase 10.1103/PhysRevE.89.022721}
  {\bibfield  {journal} {\bibinfo  {journal} {Phys. Rev. E}\ }\textbf {\bibinfo
  {volume} {89}},\ \bibinfo {pages} {022721} (\bibinfo {year}
  {2014})}\BibitemShut {NoStop}%
\bibitem [{\citenamefont {Hexner}\ and\ \citenamefont {Levine}(2015)}]{hexner}%
  \BibitemOpen
  \bibfield  {author} {\bibinfo {author} {\bibfnamefont {D.}~\bibnamefont
  {Hexner}}\ and\ \bibinfo {author} {\bibfnamefont {D.}~\bibnamefont
  {Levine}},\ }\href {\doibase 10.1103/PhysRevLett.114.110602} {\bibfield
  {journal} {\bibinfo  {journal} {Phys. Rev. Lett.}\ }\textbf {\bibinfo
  {volume} {114}},\ \bibinfo {pages} {110602} (\bibinfo {year}
  {2015})}\BibitemShut {NoStop}%
\bibitem [{\citenamefont {{R. L. Jack}}\ \emph {et~al.}(2015)\citenamefont {{R.
  L. Jack}}, \citenamefont {{I. R. Thompson}},\ and\ \citenamefont
  {Sollich}}]{jack}%
  \BibitemOpen
  \bibfield  {author} {\bibinfo {author} {\bibnamefont {{R. L. Jack}}},
  \bibinfo {author} {\bibnamefont {{I. R. Thompson}}}, \ and\ \bibinfo {author}
  {\bibfnamefont {P.}~\bibnamefont {Sollich}},\ }\href {\doibase
  10.1103/PhysRevLett.114.060601} {\bibfield  {journal} {\bibinfo  {journal}
  {Phys. Rev. Lett.}\ }\textbf {\bibinfo {volume} {114}},\ \bibinfo {pages}
  {060601} (\bibinfo {year} {2015})}\BibitemShut {NoStop}%
\bibitem [{\citenamefont {Lesanovsky}\ and\ \citenamefont {{J. P.
  Garrahan}}(2014)}]{lesanovsky}%
  \BibitemOpen
  \bibfield  {author} {\bibinfo {author} {\bibfnamefont {I.}~\bibnamefont
  {Lesanovsky}}\ and\ \bibinfo {author} {\bibnamefont {{J. P. Garrahan}}},\
  }\href {\doibase 10.1103/PhysRevA.90.011603} {\bibfield  {journal} {\bibinfo
  {journal} {Phys. Rev. A}\ }\textbf {\bibinfo {volume} {90}},\ \bibinfo
  {pages} {011603} (\bibinfo {year} {2014})}\BibitemShut {NoStop}%
\bibitem [{\citenamefont {{C. De Rosa}}\ \emph {et~al.}(2015)\citenamefont {{C.
  De Rosa}}, \citenamefont {Auriemma}, \citenamefont {Diletto}, \citenamefont
  {{R. Di Girolamo}}, \citenamefont {Malafronte}, \citenamefont {Morvillo},
  \citenamefont {Zito}, \citenamefont {Rusciano}, \citenamefont {Pesce},\ and\
  \citenamefont {Sasso}}]{ramanspectroscopy}%
  \BibitemOpen
  \bibfield  {author} {\bibinfo {author} {\bibnamefont {{C. De Rosa}}},
  \bibinfo {author} {\bibfnamefont {F.}~\bibnamefont {Auriemma}}, \bibinfo
  {author} {\bibfnamefont {C.}~\bibnamefont {Diletto}}, \bibinfo {author}
  {\bibnamefont {{R. Di Girolamo}}}, \bibinfo {author} {\bibfnamefont
  {A.}~\bibnamefont {Malafronte}}, \bibinfo {author} {\bibfnamefont
  {P.}~\bibnamefont {Morvillo}}, \bibinfo {author} {\bibfnamefont
  {G.}~\bibnamefont {Zito}}, \bibinfo {author} {\bibfnamefont {G.}~\bibnamefont
  {Rusciano}}, \bibinfo {author} {\bibfnamefont {G.}~\bibnamefont {Pesce}}, \
  and\ \bibinfo {author} {\bibfnamefont {A.}~\bibnamefont {Sasso}},\ }\href
  {\doibase 10.1039/c4cp06024e} {\bibfield  {journal} {\bibinfo  {journal}
  {Phys. Chem. Chem. Phys.}\ }\textbf {\bibinfo {volume} {17}},\ \bibinfo
  {pages} {8061} (\bibinfo {year} {2015})}\BibitemShut {NoStop}%
\bibitem [{\citenamefont {{R. Degl'Innocenti}}\ \emph
  {et~al.}(2015)\citenamefont {{R. Degl'Innocenti}}, \citenamefont {{Y. D.
  Shah}}, \citenamefont {Masini}, \citenamefont {Ronzani}, \citenamefont
  {Pitanti}, \citenamefont {Ren}, \citenamefont {{D. S. Jessop}}, \citenamefont
  {Tredicucci}, \citenamefont {{H. E. Beere}},\ and\ \citenamefont {{D. A.
  Ritchie}}}]{quantumcascadelaser}%
  \BibitemOpen
  \bibfield  {author} {\bibinfo {author} {\bibnamefont {{R. Degl'Innocenti}}},
  \bibinfo {author} {\bibnamefont {{Y. D. Shah}}}, \bibinfo {author}
  {\bibfnamefont {L.}~\bibnamefont {Masini}}, \bibinfo {author} {\bibfnamefont
  {A.}~\bibnamefont {Ronzani}}, \bibinfo {author} {\bibfnamefont
  {A.}~\bibnamefont {Pitanti}}, \bibinfo {author} {\bibfnamefont
  {Y.}~\bibnamefont {Ren}}, \bibinfo {author} {\bibnamefont {{D. S. Jessop}}},
  \bibinfo {author} {\bibfnamefont {A.}~\bibnamefont {Tredicucci}}, \bibinfo
  {author} {\bibnamefont {{H. E. Beere}}}, \ and\ \bibinfo {author}
  {\bibnamefont {{D. A. Ritchie}}},\ }\href {\doibase 10.1117/12.2083678}
  {\bibfield  {journal} {\bibinfo  {journal} {Proc. SPIE}\ }\textbf {\bibinfo
  {volume} {9370}},\ \bibinfo {pages} {93700A} (\bibinfo {year}
  {2015})}\BibitemShut {NoStop}%
\bibitem [{\citenamefont {Yu}\ \emph {et~al.}(2015)\citenamefont {Yu},
  \citenamefont {Piao}, \citenamefont {Hong},\ and\ \citenamefont
  {Park}}]{blochwavedynamics}%
  \BibitemOpen
  \bibfield  {author} {\bibinfo {author} {\bibfnamefont {S.}~\bibnamefont
  {Yu}}, \bibinfo {author} {\bibfnamefont {X.}~\bibnamefont {Piao}}, \bibinfo
  {author} {\bibfnamefont {J.}~\bibnamefont {Hong}}, \ and\ \bibinfo {author}
  {\bibfnamefont {N.}~\bibnamefont {Park}},\ }\href@noop {} {\  (\bibinfo
  {year} {2015})},\ \Eprint {http://arxiv.org/abs/1501.02591} {arXiv:1501.02591
  [physics.optics]} \BibitemShut {NoStop}%
\bibitem [{\citenamefont {Torquato}\ \emph {et~al.}(2008)\citenamefont
  {Torquato}, \citenamefont {Scardicchio},\ and\ \citenamefont {{C. E.
  Zachary}}}]{pointprocesses}%
  \BibitemOpen
  \bibfield  {author} {\bibinfo {author} {\bibfnamefont {S.}~\bibnamefont
  {Torquato}}, \bibinfo {author} {\bibfnamefont {A.}~\bibnamefont
  {Scardicchio}}, \ and\ \bibinfo {author} {\bibnamefont {{C. E. Zachary}}},\
  }\href {\doibase 10.1088/1742-5468/2008/11/P11019} {\bibfield  {journal}
  {\bibinfo  {journal} {J. Stat. Mech. Theor. Exp.}\ }\textbf {\bibinfo
  {volume} {11}},\ \bibinfo {pages} {19} (\bibinfo {year} {2008})}\BibitemShut
  {NoStop}%
\bibitem [{\citenamefont {{R. D. Batten}}\ \emph {et~al.}(2008)\citenamefont
  {{R. D. Batten}}, \citenamefont {{F. H. Stillinger}},\ and\ \citenamefont
  {Torquato}}]{stealthydisorder}%
  \BibitemOpen
  \bibfield  {author} {\bibinfo {author} {\bibnamefont {{R. D. Batten}}},
  \bibinfo {author} {\bibnamefont {{F. H. Stillinger}}}, \ and\ \bibinfo
  {author} {\bibfnamefont {S.}~\bibnamefont {Torquato}},\ }\href {\doibase
  10.1063/1.2961314} {\bibfield  {journal} {\bibinfo  {journal} {J. Appl.
  Phys.}\ }\textbf {\bibinfo {volume} {104}},\ \bibinfo {pages} {033504}
  (\bibinfo {year} {2008})}\BibitemShut {NoStop}%
\bibitem [{\citenamefont {Torquato}\ \emph {et~al.}(2015)\citenamefont
  {Torquato}, \citenamefont {Zhang},\ and\ \citenamefont {{F. H.
  Stillinger}}}]{torquato_arxiv}%
  \BibitemOpen
  \bibfield  {author} {\bibinfo {author} {\bibfnamefont {S.}~\bibnamefont
  {Torquato}}, \bibinfo {author} {\bibfnamefont {G.}~\bibnamefont {Zhang}}, \
  and\ \bibinfo {author} {\bibnamefont {{F. H. Stillinger}}},\ }\href {\doibase
  10.1103/PhysRevX.5.021020} {\bibfield  {journal} {\bibinfo  {journal} {Phys.
  Rev. X}\ }\textbf {\bibinfo {volume} {5}},\ \bibinfo {pages} {021020}
  (\bibinfo {year} {2015})}\BibitemShut {NoStop}%
\bibitem [{\citenamefont {Florescu}\ \emph {et~al.}(2009)\citenamefont
  {Florescu}, \citenamefont {Torquato},\ and\ \citenamefont {{P. J.
  Steinhardt}}}]{2009_torquato}%
  \BibitemOpen
  \bibfield  {author} {\bibinfo {author} {\bibfnamefont {M.}~\bibnamefont
  {Florescu}}, \bibinfo {author} {\bibfnamefont {S.}~\bibnamefont {Torquato}},
  \ and\ \bibinfo {author} {\bibnamefont {{P. J. Steinhardt}}},\ }\href
  {\doibase 10.1073/pnas.0907744106} {\bibfield  {journal} {\bibinfo  {journal}
  {Proc. Natl. Acad. Sci. USA}\ }\textbf {\bibinfo {volume} {106}},\ \bibinfo
  {pages} {20658} (\bibinfo {year} {2009})}\BibitemShut {NoStop}%
\bibitem [{\citenamefont {Man}\ \emph {et~al.}(2013)\citenamefont {Man},
  \citenamefont {Florescu}, \citenamefont {{E. P. Williamson}}, \citenamefont
  {He}, \citenamefont {{S. R. Hashemizad}},\ and\ \citenamefont {{B. Y. C.
  Leung}}}]{isotropicbandgaps}%
  \BibitemOpen
  \bibfield  {author} {\bibinfo {author} {\bibfnamefont {W.}~\bibnamefont
  {Man}}, \bibinfo {author} {\bibfnamefont {M.}~\bibnamefont {Florescu}},
  \bibinfo {author} {\bibnamefont {{E. P. Williamson}}}, \bibinfo {author}
  {\bibfnamefont {Y.}~\bibnamefont {He}}, \bibinfo {author} {\bibnamefont {{S.
  R. Hashemizad}}}, \ and\ \bibinfo {author} {\bibnamefont {{B. Y. C.
  Leung}}},\ }\href {\doibase 10.1073/pnas.1307879110} {\bibfield  {journal}
  {\bibinfo  {journal} {Proc. Natl. Acad. Sci. USA}\ }\textbf {\bibinfo
  {volume} {110}},\ \bibinfo {pages} {15886} (\bibinfo {year}
  {2013})}\BibitemShut {NoStop}%
\bibitem [{\citenamefont {Binder}\ and\ \citenamefont {{A. P.
  Young}}(1986)}]{spinglasses}%
  \BibitemOpen
  \bibfield  {author} {\bibinfo {author} {\bibfnamefont {K.}~\bibnamefont
  {Binder}}\ and\ \bibinfo {author} {\bibnamefont {{A. P. Young}}},\ }\href
  {\doibase 10.1103/RevModPhys.58.801} {\bibfield  {journal} {\bibinfo
  {journal} {Rev. Mod. Phys.}\ }\textbf {\bibinfo {volume} {58}},\ \bibinfo
  {pages} {801} (\bibinfo {year} {1986})}\BibitemShut {NoStop}%
\bibitem [{\citenamefont {Mezard}\ \emph {et~al.}(1987)\citenamefont {Mezard},
  \citenamefont {Parisi},\ and\ \citenamefont {Virasoro}}]{parisi}%
  \BibitemOpen
  \bibfield  {author} {\bibinfo {author} {\bibfnamefont {M.}~\bibnamefont
  {Mezard}}, \bibinfo {author} {\bibfnamefont {G.}~\bibnamefont {Parisi}}, \
  and\ \bibinfo {author} {\bibfnamefont {M.}~\bibnamefont {Virasoro}},\ }\href
  {\doibase 10.1142/0271} {\emph {\bibinfo {title} {Spin Glass Theory and
  Beyond}}}\ (\bibinfo  {publisher} {World Scientific Publishing Co. Pte. Ltd.,
  Singapore},\ \bibinfo {year} {1987})\BibitemShut {NoStop}%
\bibitem [{\citenamefont {Nishimori}(2001)}]{nishimori}%
  \BibitemOpen
  \bibfield  {author} {\bibinfo {author} {\bibfnamefont {H.}~\bibnamefont
  {Nishimori}},\ }\href {\doibase 10.1093/acprof:oso/9780198509417.001.0001}
  {\emph {\bibinfo {title} {Statistical Physics of Spin Glasses and Information
  Processing: An Introduction}}}\ (\bibinfo  {publisher} {Oxford University
  Press, Oxford},\ \bibinfo {year} {2001})\BibitemShut {NoStop}%
\bibitem [{\citenamefont {{R. J. Elliott}}(1961)}]{introANNNImodel}%
  \BibitemOpen
  \bibfield  {author} {\bibinfo {author} {\bibnamefont {{R. J. Elliott}}},\
  }\href {\doibase 10.1103/PhysRev.124.346} {\bibfield  {journal} {\bibinfo
  {journal} {Phys. Rev.}\ }\textbf {\bibinfo {volume} {124}},\ \bibinfo {pages}
  {346} (\bibinfo {year} {1961})}\BibitemShut {NoStop}%
\bibitem [{\citenamefont {{M. E. Fisher}}\ and\ \citenamefont
  {Selke}(1980)}]{ANNNImodel}%
  \BibitemOpen
  \bibfield  {author} {\bibinfo {author} {\bibnamefont {{M. E. Fisher}}}\ and\
  \bibinfo {author} {\bibfnamefont {W.}~\bibnamefont {Selke}},\ }\href
  {\doibase 10.1103/PhysRevLett.44.1502} {\bibfield  {journal} {\bibinfo
  {journal} {Phys. Rev. Lett.}\ }\textbf {\bibinfo {volume} {44}},\ \bibinfo
  {pages} {1502} (\bibinfo {year} {1980})}\BibitemShut {NoStop}%
\bibitem [{\citenamefont {Selke}(1988)}]{reviewANNNImodel}%
  \BibitemOpen
  \bibfield  {author} {\bibinfo {author} {\bibfnamefont {W.}~\bibnamefont
  {Selke}},\ }\href {\doibase 10.1016/0370-1573(88)90140-8} {\bibfield
  {journal} {\bibinfo  {journal} {Phys. Rep.}\ }\textbf {\bibinfo {volume}
  {170}},\ \bibinfo {pages} {213} (\bibinfo {year} {1988})}\BibitemShut
  {NoStop}%
\bibitem [{\citenamefont {{F. J. Dyson}}(1969)}]{dysonSpinChain}%
  \BibitemOpen
  \bibfield  {author} {\bibinfo {author} {\bibnamefont {{F. J. Dyson}}},\
  }\href {\doibase 10.1007/BF01645907} {\bibfield  {journal} {\bibinfo
  {journal} {Commun. Math. Phys.}\ }\textbf {\bibinfo {volume} {12}},\ \bibinfo
  {pages} {91} (\bibinfo {year} {1969})}\BibitemShut {NoStop}%
\bibitem [{\citenamefont {Kotliar}\ \emph {et~al.}(1983)\citenamefont
  {Kotliar}, \citenamefont {{P. W. Anderson}},\ and\ \citenamefont {{D. L.
  Stein}}}]{1Dspinglass}%
  \BibitemOpen
  \bibfield  {author} {\bibinfo {author} {\bibfnamefont {G.}~\bibnamefont
  {Kotliar}}, \bibinfo {author} {\bibnamefont {{P. W. Anderson}}}, \ and\
  \bibinfo {author} {\bibnamefont {{D. L. Stein}}},\ }\href {\doibase
  10.1103/PhysRevB.27.602} {\bibfield  {journal} {\bibinfo  {journal} {Phys.
  Rev. B}\ }\textbf {\bibinfo {volume} {27}},\ \bibinfo {pages} {602} (\bibinfo
  {year} {1983})}\BibitemShut {NoStop}%
\bibitem [{\citenamefont {Luijten}\ and\ \citenamefont {{H. W. J.
  Bl\"ote}}(1997)}]{lujiten}%
  \BibitemOpen
  \bibfield  {author} {\bibinfo {author} {\bibfnamefont {E.}~\bibnamefont
  {Luijten}}\ and\ \bibinfo {author} {\bibnamefont {{H. W. J. Bl\"ote}}},\
  }\href {\doibase 10.1103/PhysRevB.56.8945} {\bibfield  {journal} {\bibinfo
  {journal} {Phys. Rev. B}\ }\textbf {\bibinfo {volume} {56}},\ \bibinfo
  {pages} {8945} (\bibinfo {year} {1997})}\BibitemShut {NoStop}%
\bibitem [{\citenamefont {{M. C. Angelini}}\ \emph {et~al.}(2014)\citenamefont
  {{M. C. Angelini}}, \citenamefont {{G. Parisi}},\ and\ \citenamefont {{F.
  Ricci-Tersenghi}}}]{parisiLRspinchain}%
  \BibitemOpen
  \bibfield  {author} {\bibinfo {author} {\bibnamefont {{M. C. Angelini}}},
  \bibinfo {author} {\bibnamefont {{G. Parisi}}}, \ and\ \bibinfo {author}
  {\bibnamefont {{F. Ricci-Tersenghi}}},\ }\href {\doibase
  10.1103/PhysRevE.89.062120} {\bibfield  {journal} {\bibinfo  {journal} {Phys.
  Rev. E}\ }\textbf {\bibinfo {volume} {89}},\ \bibinfo {pages} {062120}
  (\bibinfo {year} {2014})}\BibitemShut {NoStop}%
\bibitem [{\citenamefont {{J.-P. Hansen}}\ and\ \citenamefont {{I. R.
  McDonald}}(1986)}]{simpleliquids}%
  \BibitemOpen
  \bibfield  {author} {\bibinfo {author} {\bibnamefont {{J.-P. Hansen}}}\ and\
  \bibinfo {author} {\bibnamefont {{I. R. McDonald}}},\ }\href {\doibase
  10.1016/B978-0-12-387032-2.00013-1} {\emph {\bibinfo {title} {Theory of
  Simple Liquids}}}\ (\bibinfo  {publisher} {Academic Press, New York},\
  \bibinfo {year} {1986})\BibitemShut {NoStop}%
\bibitem [{\citenamefont {{R. A. DiStasio Jr.}}\ \emph {et~al.}()\citenamefont
  {{R. A. DiStasio Jr.}}, \citenamefont {Zhang}, \citenamefont {{F. H.
  Stillinger}},\ and\ \citenamefont {Torquato}}]{robpaper}%
  \BibitemOpen
  \bibfield  {author} {\bibinfo {author} {\bibnamefont {{R. A. DiStasio Jr.}}},
  \bibinfo {author} {\bibfnamefont {G.}~\bibnamefont {Zhang}}, \bibinfo
  {author} {\bibnamefont {{F. H. Stillinger}}}, \ and\ \bibinfo {author}
  {\bibfnamefont {S.}~\bibnamefont {Torquato}},\ }\href@noop {} {\enquote
  {\bibinfo {title} {Rational design of stealthy hyperuniform patterns with
  tunable order},}\ }\bibinfo {note} {(to be published)}\BibitemShut {NoStop}%
\bibitem [{\citenamefont {{Marinari}}\ \emph {et~al.}(1994)\citenamefont
  {{Marinari}}, \citenamefont {{Parisi}},\ and\ \citenamefont
  {{Ritort}}}]{parisibinary}%
  \BibitemOpen
  \bibfield  {author} {\bibinfo {author} {\bibfnamefont {E.}~\bibnamefont
  {{Marinari}}}, \bibinfo {author} {\bibfnamefont {G.}~\bibnamefont
  {{Parisi}}}, \ and\ \bibinfo {author} {\bibfnamefont {F.}~\bibnamefont
  {{Ritort}}},\ }\href {\doibase 10.1088/0305-4470/27/23/010} {\bibfield
  {journal} {\bibinfo  {journal} {J. Phys. A}\ }\textbf {\bibinfo {volume}
  {27}},\ \bibinfo {pages} {7615} (\bibinfo {year} {1994})}\BibitemShut
  {NoStop}%
\bibitem [{\citenamefont {Yeong}\ and\ \citenamefont
  {Torquato}(1998)}]{reconstruction}%
  \BibitemOpen
  \bibfield  {author} {\bibinfo {author} {\bibfnamefont {C.~L.~Y.}\
  \bibnamefont {Yeong}}\ and\ \bibinfo {author} {\bibfnamefont
  {S.}~\bibnamefont {Torquato}},\ }\href {\doibase 10.1103/PhysRevE.57.495}
  {\bibfield  {journal} {\bibinfo  {journal} {Phys. Rev. E}\ }\textbf {\bibinfo
  {volume} {57}},\ \bibinfo {pages} {495} (\bibinfo {year} {1998})}\BibitemShut
  {NoStop}%
\bibitem [{\citenamefont {{O. U. Uche}}\ \emph {et~al.}(2004)\citenamefont {{O.
  U. Uche}}, \citenamefont {{F. H. Stillinger}},\ and\ \citenamefont
  {Torquato}}]{stealthy2}%
  \BibitemOpen
  \bibfield  {author} {\bibinfo {author} {\bibnamefont {{O. U. Uche}}},
  \bibinfo {author} {\bibnamefont {{F. H. Stillinger}}}, \ and\ \bibinfo
  {author} {\bibfnamefont {S.}~\bibnamefont {Torquato}},\ }\href {\doibase
  10.1103/PhysRevE.70.046122} {\bibfield  {journal} {\bibinfo  {journal} {Phys.
  Rev. E}\ }\textbf {\bibinfo {volume} {70}},\ \bibinfo {pages} {046122}
  (\bibinfo {year} {2004})}\BibitemShut {NoStop}%
\end{thebibliography}%

\end{document}